\documentclass[prd,twocolumn,amsmath,amssymb,nofootinbib,floatfix,superscriptaddress]{revtex4}

\usepackage{graphicx,bm,float}

\makeatletter
\def\graphicscale{\twocolumn@sw{0.3}{0.4}}
\def\graphicthreescale{\twocolumn@sw{0.3}{0.4}}

\begin{document}

\title{Universal low-temperature behavior of 
two-dimensional lattice scalar chromodynamics}

\author{Claudio Bonati} 
\affiliation{Dipartimento di Fisica dell'Universit\`a di Pisa 
       and INFN Largo Pontecorvo 3, I-56127 Pisa, Italy}

\author{Andrea Pelissetto}
\affiliation{Dipartimento di Fisica dell'Universit\`a di Roma Sapienza
        and INFN Sezione di Roma I, I-00185 Roma, Italy}

\author{Ettore Vicari} 
\affiliation{Dipartimento di Fisica dell'Universit\`a di Pisa
       and INFN Largo Pontecorvo 3, I-56127 Pisa, Italy}

\date{\today}

\begin{abstract}
We study the role that global and local nonabelian symmetries play in
two-dimensional lattice gauge theories with multicomponent scalar
fields.  We start from a maximally O($M$)-symmetric multicomponent
scalar model, Its symmetry is partially gauged to obtain an SU($N_c$)
gauge theory (scalar chromodynamics) with global U$(N_f)$ (for $N_c\ge
3$) or Sp($N_f$) symmetry (for $N_c=2$), where $N_f>1$ is the number
of flavors. Correspondingly, the fields belong to the coset
$S^M$/SU($N_c$) where $S^M$ is the $M$-dimensional sphere and $M=2 N_f
N_c$.  In agreement with the Mermin-Wagner theorem, the system is
always disordered at finite temperature and a critical behavior only
develops in the zero-temperature limit.  Its universal features are
investigated by numerical finite-size scaling methods.  The results
show that the asymptotic low-temperature behavior belongs to the
universality class of the 2D CP$^{N_f-1}$ field theory for $N_c>2$,
and to that of the 2D Sp($N_f$) field theory for $N_c=2$. These
universality classes correspond to 2D statistical field theories
associated with symmetric spaces that are invariant under Sp($N_f$)
transformations for $N_c=2$ and under SU($N_f$) for $N_c > 2$. These
symmetry groups are the same invariance groups of scalar
chromodynamics, apart from a U(1) flavor symmetry that is present for
$N_f \ge N_c > 2$, which does not play any role in determining the
asymptotic behavior of the model.
\end{abstract}

\maketitle


\section{Introduction}
\label{intro}

Nonabelian gauge symmetries are known since long time to describe
fundamental interactions~\cite{Weinberg-book}. More recently, it has
been pointed out that they may also characterize emerging phenomena in
condensed-matter physics, see, e.g.,
Refs.~\cite{WNMXS-17,GASVW-18,SSST-19,GSF-19,Sachdev-19} and
references therein.  As a consequence, the large-scale properties of
gauge models are also of interest in two or three dimensions.

We consider a lattice model of interacting scalar fields in the
presence of nonabelian gauge symmetries, which may be named scalar
chromodynamics or nonabelian Higgs model. In four space-time
dimensions it represents a paradigmatic example to discuss the
nonabelian Higgs mechanism, which is at the basis of the Standard
Model of fundamental interactions.  The three-dimensional model may
also be relevant in condensed-matter physics, for systems with
emerging nonabelian gauge symmetries. Its phase diagram and its
behavior at the finite-temperature phase transitions has been
investigated in Refs.~\cite{BPV-19,BPV-20}. In this paper we extend
such a study to two-dimensional (2D) systems.

We consider a 2D lattice nonabelian gauge theory with multicomponent
scalar fields.  It is defined starting from a maximally
O($M$)-symmetric multicomponent scalar model. The global symmetry is
partially gauged, obtaining a nonabelian gauge model, in which the
fields belong to the coset $S^M$/SU($N_c$), where $M=2 N_f N_c$,
$N_f$
is the number of flavors, and $S^M=\hbox{SO}(M)/\hbox{SO}(M-1)$ is the
$M$-dimensional sphere.  According to the Mermin-Wagner
theorem~\cite{MW-66}, the model is always disordered for finite values
of the temperature.  However, a critical behavior develops in the
zero-temperature limit. We investigate its universal features for
generic values of $N_c$ and $N_f\ge 2$, by means of finite-size
scaling (FSS) analyses of Monte Carlo (MC) simulations.

The results provide numerical evidence that the asymptotic
low-temperature behavior of these lattice nonabelian gauge models
belongs to the universality class of the 2D CP$^{N_f-1}$ field theory
when $N_c\ge 3$, and to that of the 2D Sp($N_f$) field theory for
$N_c=2$. This suggests that the renormalization-group (RG) flow of the
2D multiflavor lattice scalar chromodynamics associated with the coset
$S^M$/SU($N_c$) is asymptotically controlled by the 2D statistical
field theories associated with the symmetric
spaces~\cite{BHZ-80,ZJ-book} that have the same global symmetry, i.e.,
SU($N_f$) for $N_c\ge 3$ and Sp($N_f$) for $N_c=2$.

The paper is organized as follows. In Sec.~\ref{SQCD} we introduce the
lattice nonabelian gauge models that we consider. In Sec.~\ref{fsssec}
we discuss the general strategy we use to investigate the nature of
the low-temperature critical behavior.  Then, in
Secs.~\ref{resultsncl3} and \ref{resnc2} we report the numerical
results for lattice models with $N_c\ge 3$ and $N_c=2$, respectively.
Finally, in Sec.~\ref{conclu} we summarize and draw our conclusions.
In App.~\ref{largebeta} we report some results on the minimum-energy
configurations of the models considered.

\section{Multiflavor lattice scalar chromodynamics}
\label{SQCD}

We consider a 2D lattice scalar nonabelian gauge theory obtained by
partially gauging a maximally symmetric model of complex matrix
variables $\varphi_{\bm x}^{af}$, where the indices $a=1,..,N_c$ and
$f=1,...,N_f$ are associated with the color and flavor degrees of
freedom, respectively.

We start from the maximally symmetric action
\begin{eqnarray}
S_s = - t \sum_{{\bm x},\mu} {\rm Re} \,
{\rm Tr}\,\varphi_{\bm x}^\dagger \varphi_{{\bm x}+\hat\mu} \,,
\qquad 
{\rm  Tr}\,\varphi_{\bm x}^\dagger \varphi_{\bm x} = 1\,, 
\label{ullimit}
\end{eqnarray} 
where the sum is over all sites and links of a square lattice and
$\hat{\mu}=\hat{1},\hat{2}$ denote the unit vectors along the lattice
directions. Model (\ref{ullimit}) with the unit-length constraint for
the $\varphi_{\bm x}$ variables is a particular limit of a model with
a quartic potential $\sum_{\bm x} V( {\rm Tr}\, \varphi^\dagger_{\bm
  x}\,\varphi_{\bm x})$ of the form $V(X) = r X + {1\over 2} u\,
X^2$. Indeed, it can be obtained by simply setting $r+u=0$ and taking
the limit $u\to\infty$.  In the following we set $t=1$ for simplicity,
which amounts to an appropriate choice of the temperature unit. It is
simple to see that the action $S_s$ has a global O($M$) symmetry, with
$M=2N_f N_c$. Indeed, it can also be written in terms of $M$-component
real vectors ${\bm s}_{\bm x}$ (which are the real and imaginary parts
of $\varphi_{\bm x}^{af})$ as
\begin{eqnarray}
S_s = -  \sum_{{\bm x},\mu} {\bm s}_{\bm x}\cdot 
{\bm s}_{{\bm x}+\hat{\mu}}\,,
\qquad 
{\bm s}_{\bm x} \cdot {\bm s}_{\bm x} = 1\,.
\label{ullimits}
\end{eqnarray} 
This is the standard nearest-neighbor $M$-vector lattice model.

We proceed by gauging some of the degrees of freedom using the Wilson
approach~\cite{Wilson-74}. We associate an SU($N_c$) matrix $U_{{\bm
    x},\mu}$ with each lattice link [$({\bm x},\mu)$ denotes the link
  that starts at site ${\bm x}$ in the $\hat\mu$ direction] and add a
Wilson kinetic term for the gauge fields.  We obtain the action of the
2D lattice scalar chromodynamics defined by
\begin{eqnarray}
S_g  = - N_f \sum_{{\bm x},\mu} 
{\rm Re}\, {\rm Tr} \,\varphi_{\bm x}^\dagger \, U_{{\bm x},\mu}
\, \varphi_{{\bm x}+\hat{\mu}} 
- {\gamma\over N_c}  \sum_{{\bm x}} {\rm Re} \, {\rm Tr}\,
\Pi_{\bm x},\;
\label{hgauge}
\end{eqnarray}
where $\Pi_{\bm x}$ is the plaquette operator
\begin{equation}
\Pi_{\bm x}= 
U_{{\bm x},\hat{1}} \,U_{{\bm x}+\hat{1},2} 
\,U_{{\bm x}+\hat{2},1}^\dagger  
\,U_{{\bm x},2}^\dagger
\,.
\label{plaquette}
\end{equation}
The plaquette parameter $\gamma$ plays the role of inverse gauge
coupling, and the $N_f$ and $N_c$ factors in Eq.~\eqref{hgauge} are
conventional. The partition function reads
\begin{eqnarray}
Z = \sum_{\{\varphi,U\}} e^{-\beta \,S_g}\,,\qquad \beta\equiv 1/T\,.
\label{partfun}
\end{eqnarray}
The lattice model (\ref{hgauge}) is invariant under
SU($N_c$) gauge transformations:
\begin{equation}
\varphi_{\bm x}\to W_{\bm x} \varphi_{\bm x}\,,\qquad
U_{{\bm x},\mu} \to W_{\bm x} U_{{\bm x},\mu} W_{{\bm x} +
  \hat{\mu}}^\dagger\,,
\label{gautra}
\end{equation}  
with $W_{\bm x}\in {\rm SU}(N_c)$.  For $\gamma\to\infty$, the link
variables $U_{\bm x}$ become equal to the identity (modulo gauge
transformations), thus one recovers the ungauged model
(\ref{ullimit}), or equivalently the O($M$) vector model
(\ref{ullimits}).

For $N_f=1$ the model is trivial. Because of the unit-length
condition, using gauge transformations we can fix $\varphi_{\bm x}$ to
any given unit-length vector on the whole lattice: there is no
dynamics associated with the scalar field. As we shall see,
multiflavor models with $N_f \ge 2$ show instead a nontrivial
behavior.

After gauging, the residual global symmetry depends on the number of
flavors $N_f$ and of colors $N_c$. For $N_c\ge 3$, model
(\ref{hgauge}) is invariant under the transformation
\begin{equation}
\varphi_{\bm x} \to
\varphi_{\bm x} \, V\,,\qquad V\in {\rm U}(N_f)\,,
\label{varphitra}
\end{equation}
thus it has a global U($N_f$)$/\mathbb{Z}_{N_c}$ symmetry,
$\mathbb{Z}_{N_c}$ being the center of SU($N_c$).  As discussed in
Ref.~\cite{BPV-20}, when $N_f<N_c$
\begin{equation}
\varphi_{\bm x} \to e^{i\theta} \varphi_{\bm x}
\label{u1varphi}
\end{equation}
can be realized by an appropriate SU($N_c$) local
transformation. Thus, the actual global symmetry group reduces to
SU($N_f$).

For $N_c=2$, model (\ref{hgauge}) is invariant under the larger group
Sp($N_f$)$/\mathbb{Z}_2$, where Sp$(N_f) \supset \hbox{U}(N_f)$ is the
compact complex symplectic group, see also
Refs.~\cite{WNMXS-17,Georgi-book,DP-14,BPV-19, BPV-20}. Indeed, if one
defines the 2$\times 2N_f$ matrix field
\begin{equation}
\Gamma_{\bm x}^{af} = \varphi_{\bm x}^{af}\,, \qquad \Gamma_{\bm
  x}^{a(N_f + f)} = \sum_{b} \epsilon^{ab} \bar{\varphi}_{\bm
  x}^{bf}\,,
\label{Gammadef}
\end{equation}
where $f=1,...,N_f$, $\epsilon^{ab}=-\epsilon^{ba}$,
$\epsilon^{12}=1$, the action (\ref{hgauge}) is invariant under the
global transformation
\begin{equation}
{\Gamma}_{\bm x}^{al} \to \sum_{m=1}^{2N_f} \Gamma_{\bm x}^{am} Y^{ml}\,,
\qquad Y \in {\rm Sp}(N_f)\,.
\label{invtraspn}
\end{equation}
We recall that the compact complex symplectic group Sp$(N_f)$ is the
group of the $2N_f\times 2N_f$ unitary matrices $U_{\rm sp}$
satisfying the condition
\begin{equation}
U_{\rm sp} \, J \, U_{\rm sp}^T = J\,,\qquad 
J=
\left(\begin{array}{cc} \phantom{}0 & -I \\ I &
  \phantom{-}0\end{array}\right)\,,
\label{mscond}
\end{equation}
where $I$ is the $N_f\times N_f$ identity matrix.

\section{Universal finite-size scaling}
\label{fsssec}

We exploit FSS techniques~\cite{FB-72,Barber-83,Privman-90,PV-02} to
study the nature of the asymptotic critical behavior of the model for
$T\to 0$.  For this purpose we consider models defined on square
lattices of linear size $L$ with periodic boundary conditions.

We mostly focus on the correlations of the gauge-invariant variable
$Q_{\bm x}$ defined by
\begin{equation}
Q_{{\bm x}}^{fg} = P_{\bm x}^{fg} 
-{1\over N_f} \delta^{fg}\,,
\qquad
P_{\bm x}^{fg} = \sum_a \bar{\varphi}_{\bm x}^{af} 
\varphi_{\bm x}^{ag} \,,
\label{qdef}
\end{equation}
which is a hermitian and traceless $N_f\times N_f$ matrix.  The
corresponding two-point correlation function is defined as
\begin{equation}
G({\bm x}-{\bm y}) = \langle {\rm Tr}\, Q_{\bm x} Q_{\bm y} \rangle\,,  
\label{gxyp}
\end{equation}
where the translation invariance of the system has been taken into
account. We define the susceptibility $\chi=\sum_{\bm x} G({\bm x})$
and the correlation length
\begin{eqnarray}
\xi^2 = {1\over 4 \sin^2 (\pi/L)}
{\widetilde{G}({\bm 0}) - \widetilde{G}({\bm p}_m)\over 
\widetilde{G}({\bm p}_m)}\,,
\label{xidefpb}
\end{eqnarray}
where $\widetilde{G}({\bm p})=\sum_{{\bm x}} e^{i{\bm p}\cdot {\bm x}}
G({\bm x})$ is the Fourier transform of $G({\bm x})$, and ${\bm p}_m =
(2\pi/L,0)$. We also consider the quartic cumulant (Binder) parameter
defined as
\begin{equation}
U = {\langle \mu_2^2\rangle \over \langle \mu_2 \rangle^2} \,, \qquad
\mu_2 = {1\over V^2}
\sum_{{\bm x},{\bm y}} {\rm Tr}\,Q_{\bm x} Q_{\bm y}\,,
\label{binderdef}
\end{equation}
where $V=L^2$.

To identify the universality class of the asymptotic zero-temperature
behavior, we consider the Binder parameter $U$ as a function of the
ratio
\begin{equation}
R_\xi\equiv \xi/L\,.
\label{rxidef}
\end{equation}
Indeed, in the FSS limit we have (see, e.g., Ref.~\cite{BPV-19-2})
\begin{equation}
U(\beta,L) \approx F(R_\xi)\,,
\label{r12sca}
\end{equation}
where $F(x)$ is a universal scaling function that completely
characterizes the universality class of the transition.
Eq.~(\ref{r12sca}) is particularly convenient, as it allows us to
check the universality of the asymptotic zero-temperature behavior
without the need of tuning any parameter.  Corrections to
Eq.~(\ref{r12sca}) decay as a power of $L$. In the case of
asymptotically free models, such as the 2D CP$^{N-1}$ and O($N$)
vector models, corrections decrease as $L^{-2}$, multiplied by powers
of $\ln L$ ~\cite{BPV-19-2,CP-98}.

Because of the universality of relation (\ref{r12sca}), we can use the
plots of $U$ versus $R_\xi$ to identify the models that belong to the
same universality class. If the data of $U$ for two different models
follow the same curve when plotted versus $R_\xi$, their critical
behavior is described by the same continuum quantum field theory.
This implies that any other dimensionless RG invariant quantity has
the same critical behavior in the two models, both in the
thermodynamic and in the FSS limit.  An analogous strategy was
employed in Ref.~\cite{BPV-19-2} to study the critical behavior of the
2D Abelian-Higgs lattice model in the zero-temperature limit.

The asymptotic values of $F(R_{\xi})$ for $R_{\xi}\to 0$ and
$R_{\xi}\to\infty$ correspond to the values that $U$ takes in the
small-$\beta$ and large-$\beta$ limits.  For $R_\xi\to 0$ we have
\begin{eqnarray}
\lim_{R_\xi\to 0} \;U = {N_f^2+1\over N_f^2-1}\, .
\label{uextr}
\end{eqnarray}
independently of the value of $N_c$. The large-$\beta$ limit is
discussed in App.~\ref{largebeta}. For $N_c \ge 3$ we have simply $U =
1$.

In the following we study the large-$\beta$ critical behavior of
lattice scalar chromodynamics for several values of $N_f$ and
$N_c$. We perform numerical simulations using the same upgrading
algorithm employed in three dimensions \cite{BPV-19, BPV-20}.  The
analysis of the data of $U$ versus $R_\xi$ outlined above allows us to
conclude that the critical behavior only depends on the global
symmetry group of the model. For any $N_c \ge 3$, the critical
behavior belongs to the universality class of the 2D CP$^{N_f-1}$
field theory.  Indeed, the FSS curves (\ref{r12sca}) for the model
(\ref{hgauge}) agree with those computed in the CP$^{N-1}$ model (we
use the results reported in Ref.~\cite{BPV-19-2}). For $N_c = 2$,
instead, the critical behavior is associated with that of the 2D
Sp($N_f$) field theory. Note that the parameter $\gamma$ appears to be
irrelevant in the RG sense (at least for $|\gamma|$ not too
large). Indeed, for all positive and negative values of $N_c$, $N_f$,
and $\gamma$ we investigated, the universal critical behavior does not
depend on $\gamma$.

\section{SU($N_c$) gauge models with $N_c\ge 3$}
\label{resultsncl3}

\subsection{Numerical results}
\label{resnf2ncl2}

In this section we study the critical behavior of scalar
chromodynamics for some values of $N_f$ and of $N_c\ge 3$. We compute
the scaling curve (\ref{r12sca}) and compare it with the corresponding
one computed in the CP$^{N_f-1}$ model. Such a comparison provides
evidence that the asymptotic zero-temperature behavior for finite
values of $\gamma$ in a wide interval around $\gamma=0$ is described
by the 2D CP$^{N_f-1}$ field theory.

\begin{figure}[tbp]
\includegraphics*[scale=\graphicscale]{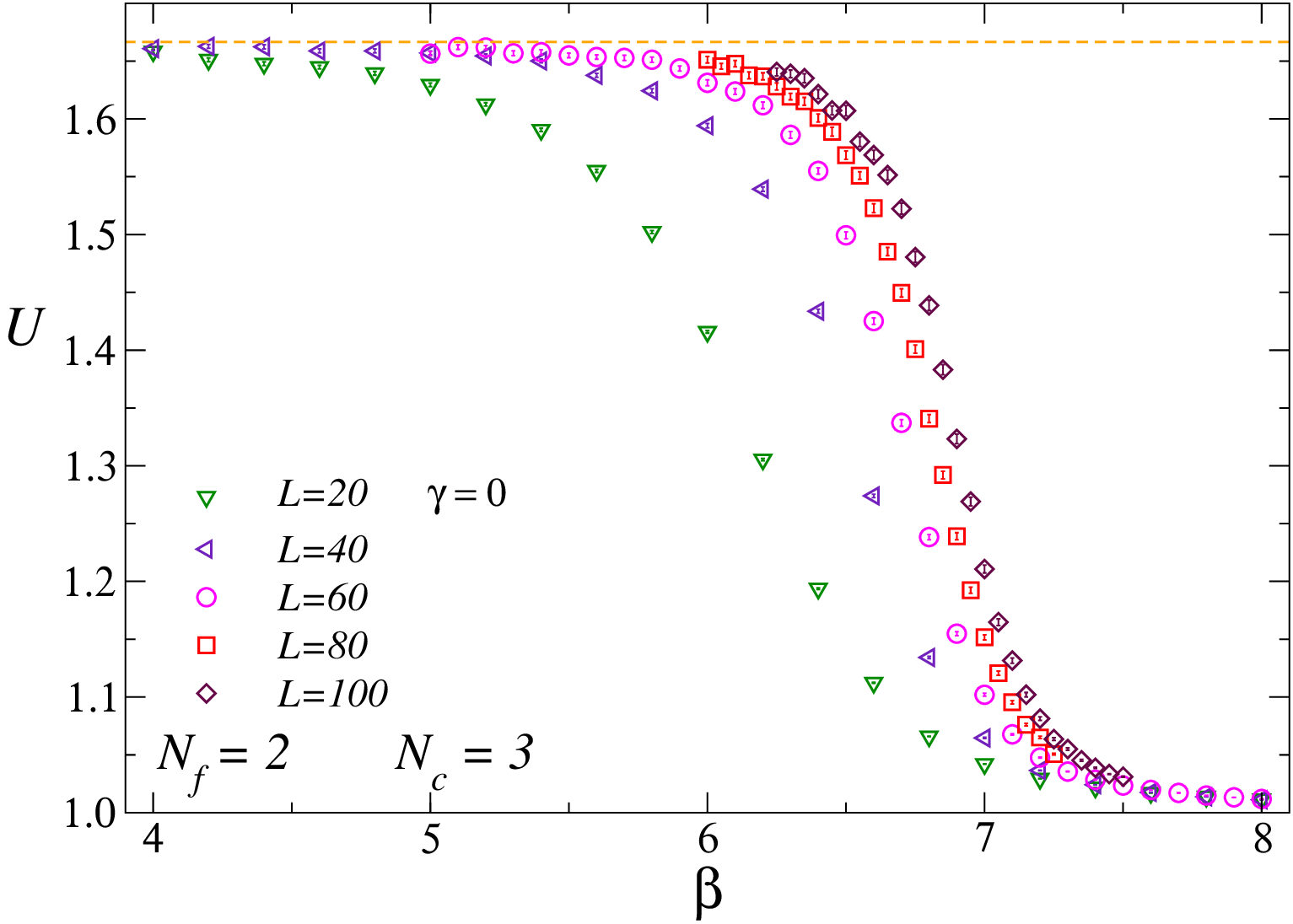}
\caption{Plot of $U$ versus $\beta$ for $N_f=2$, $N_c=3$, and
  $\gamma=0$.  The horizontal dashed line corresponds to $U=5/3$, the
  asymptotic value for $\beta \to 0$.  }
\label{u-beta-nf2nc3}
\end{figure}

\begin{figure}[tbp]
\includegraphics*[scale=\graphicscale]{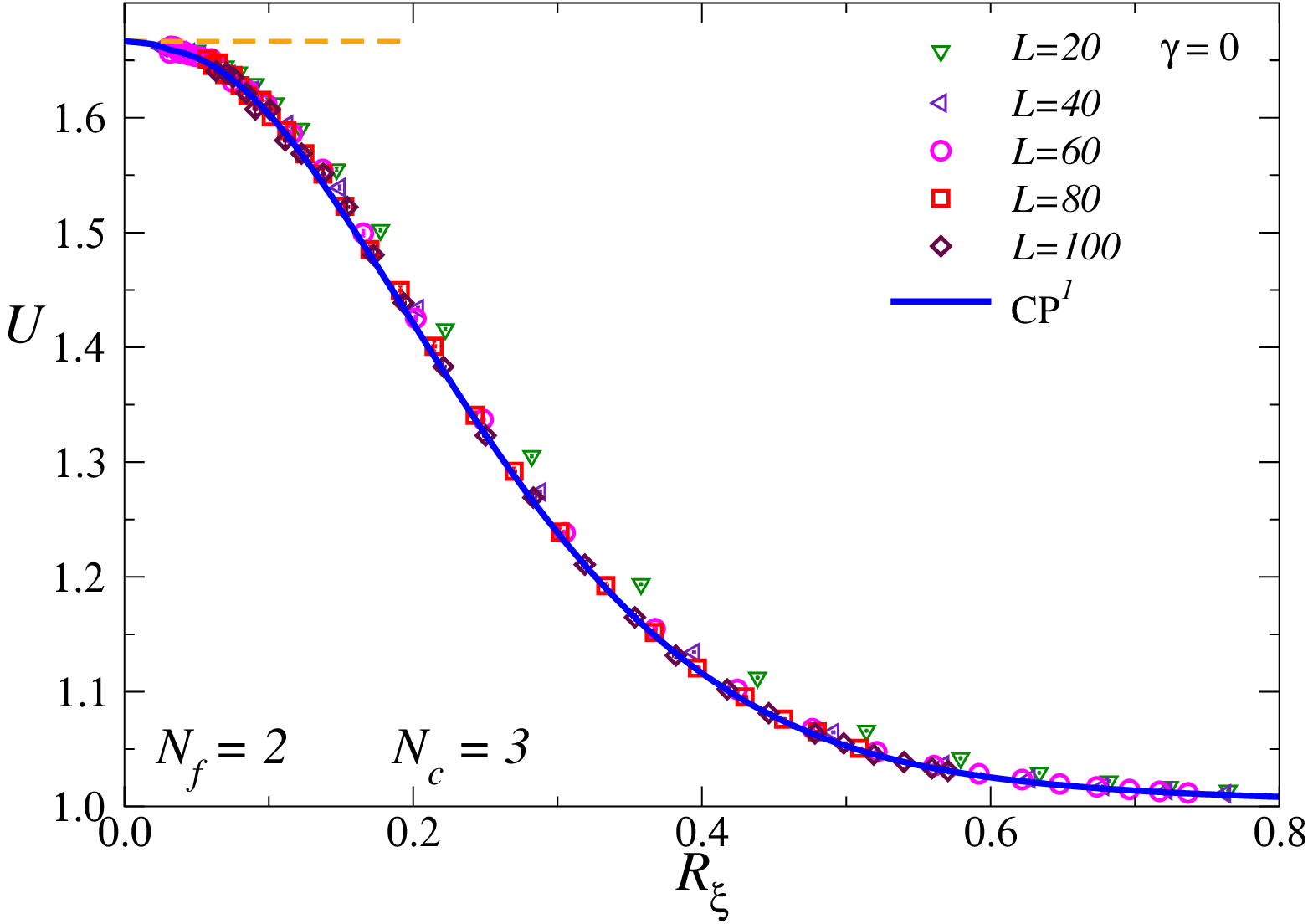}
\caption{Plot of $U$ versus $R_\xi$ for $N_f=2$, $N_c=3$, and
  $\gamma=0$.  Data approach the universal FSS curve of the 2D CP$^1$
  or O(3) universality class (full line, taken from
  Ref.~\cite{BPV-19-2}).  The horizontal dashed line corresponds to
  $U=5/3$, the asymptotic value for $R_\xi\to 0$. }
\label{u-rxi-nf2nc3}
\end{figure}

In Figs.~\ref{u-beta-nf2nc3} and \ref{u-rxi-nf2nc3} we show MC data
for the two-flavor model (\ref{hgauge}) with SU(3) gauge symmetry,
i.e., for $N_f=2$ and $N_c=3$, and $\gamma=0$.  In
Fig.~\ref{u-beta-nf2nc3} the results for the Binder parameter are
shown as a function of $\beta$ for several lattice sizes. The curves
corresponding to different lattice sizes do not intersect, confirming
the absence of a phase transition at finite $\beta$, as expected from
the Mermin-Wagner theorem. The ratio $R_\xi$ behaves analogously.  For
each lattice size $R_{\xi}$ is an increasing function of $\beta$,
seemingly divergent for $\beta\to\infty$, but no crossing is present
between curves corresponding to different $L$ values.  In
Fig.~\ref{u-rxi-nf2nc3} the data of $U$ appear to approach a FSS curve
in the large-$L$ limit when plotted versus $R_\xi$, in agreement with
the FSS prediction (\ref{r12sca}).  This asymptotic FSS curve is
consistent with that of the 2D CP$^1$ universality class (equivalent
to that of the O(3) vector model , see, e.g., Ref.~\cite{ZJ-book})
determined in Ref.~\cite{BPV-19-2}.  Moreover, scaling corrections are
consistent with the expected $O(L^{-2})$ behavior.

\begin{figure}[tbp]
\includegraphics*[scale=\graphicscale]{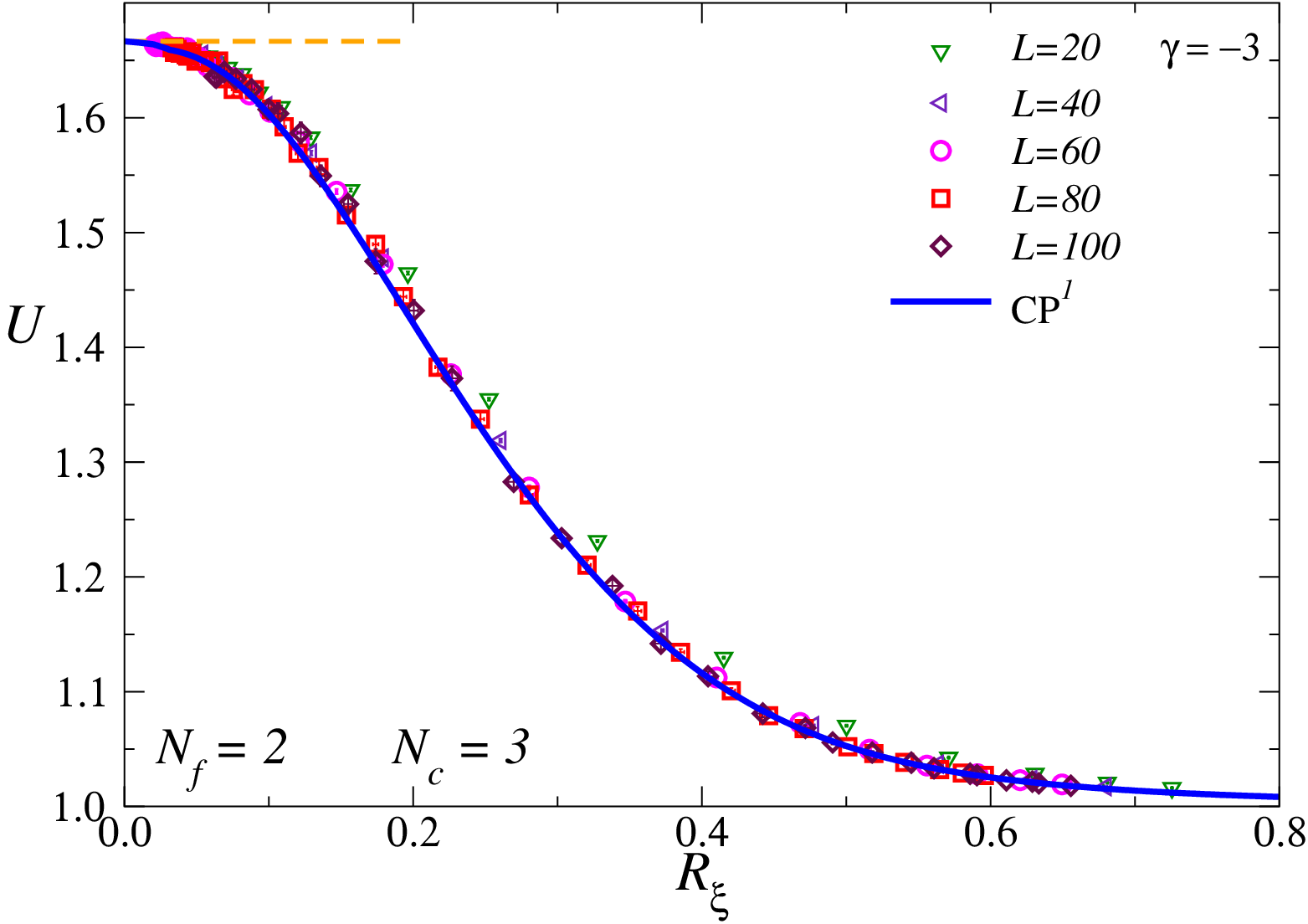}
\includegraphics*[scale=\graphicscale]{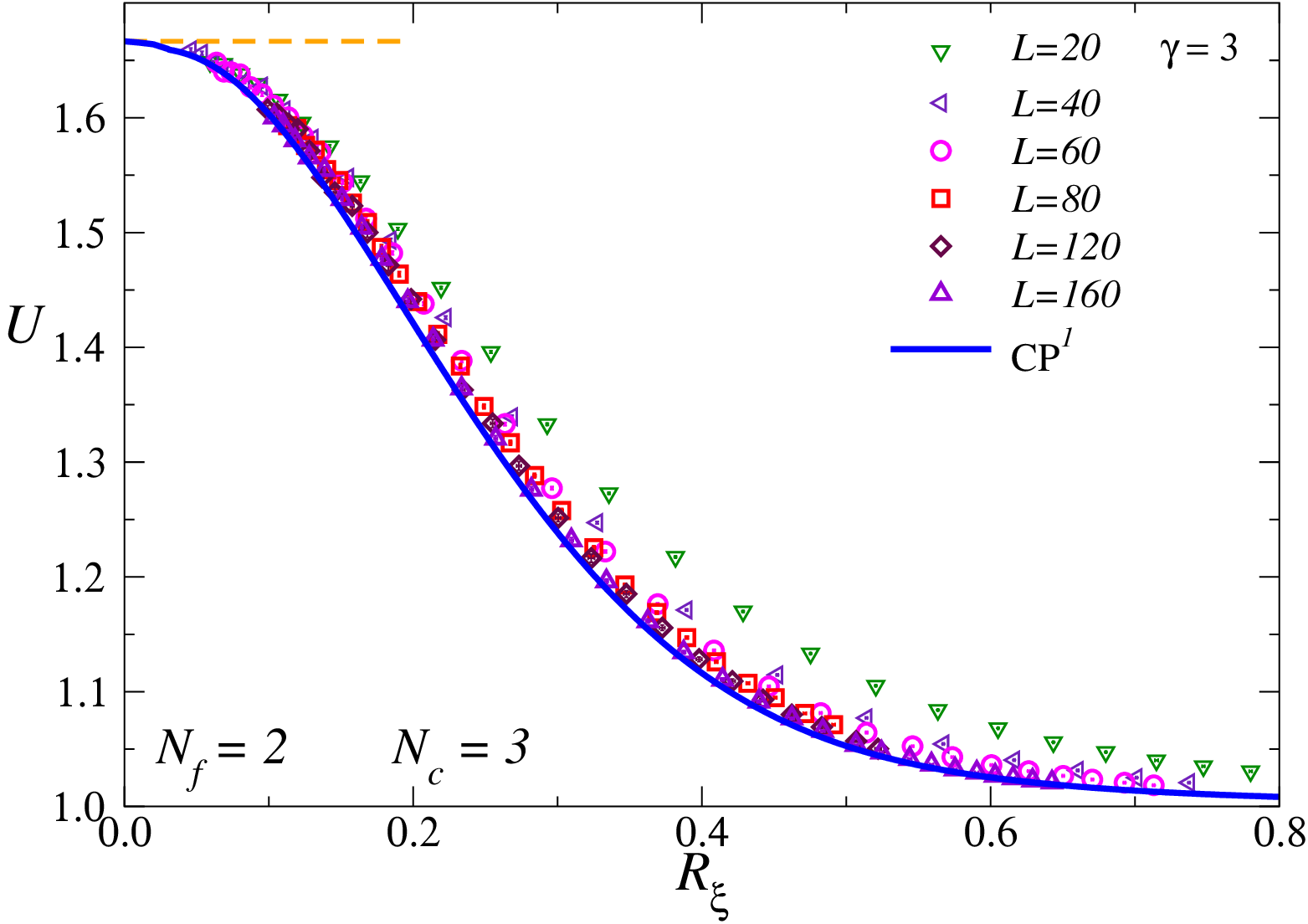}
\caption{Plot of $U$ versus $R_\xi$ for $N_f=2$, $N_c=3$, and
  $\gamma=-3$ (upper panel) and $\gamma=3$ (lower panel). Data
  approach the universal FSS curve of the 2D CP$^1$ or O(3)
  universality class (full line, taken from Ref.~\cite{BPV-19-2}).
  The horizontal dashed line corresponds to $U=5/3$, the asymptotic
  value for $R_\xi\to 0$. }
\label{u-rxi-nf2nc3-gamma}
\end{figure}

\begin{figure}[tbp]
\includegraphics*[scale=\graphicscale]{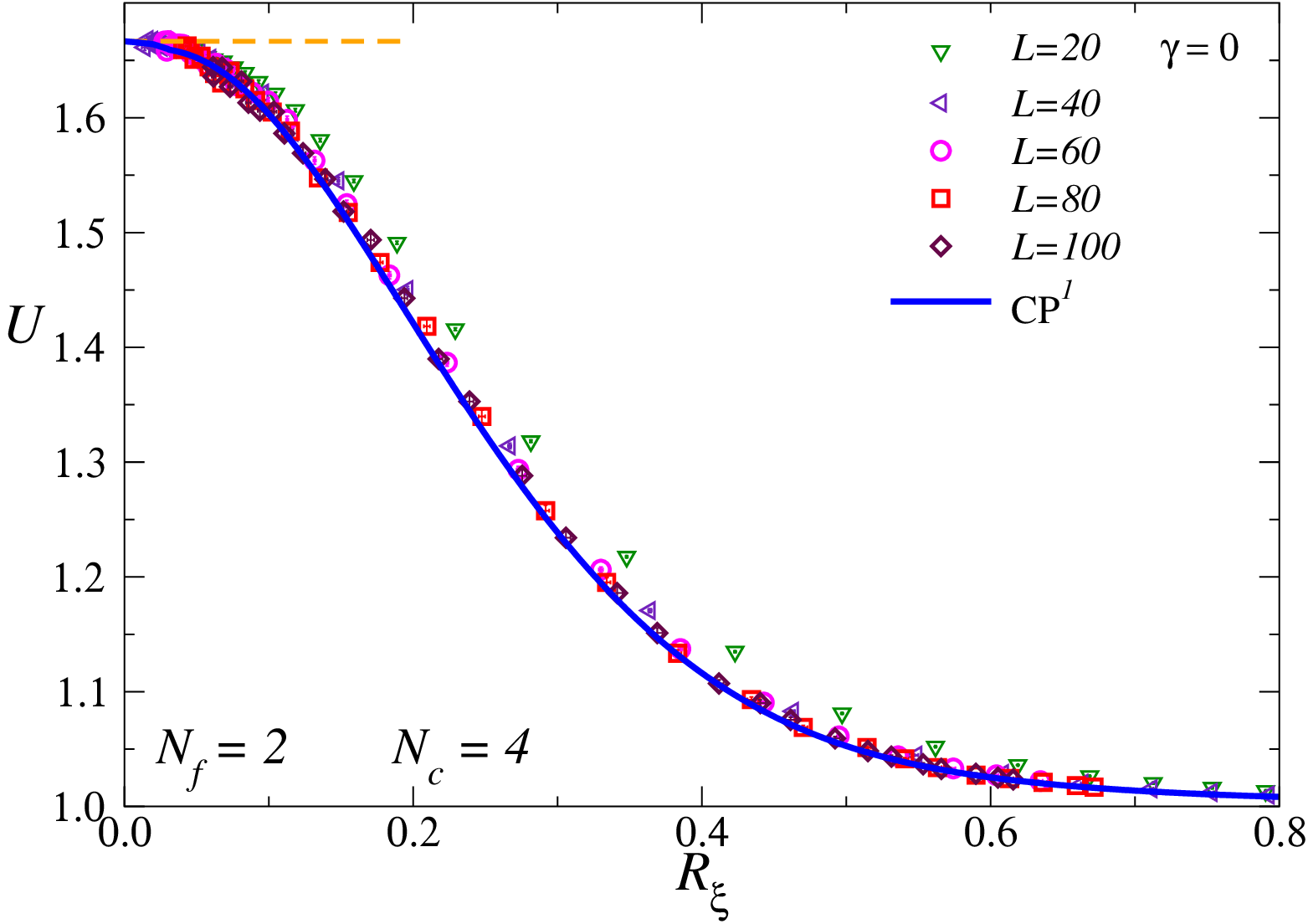}
\caption{Plot of $U$ versus $R_\xi$ for $N_f=2$, $N_c=4$, and
  $\gamma=0$. Data approach the universal FSS curve of the 2D CP$^1$
  or O(3) universality class (full line, taken from
  Ref.~\cite{BPV-19-2}).  The horizontal dashed line corresponds to
  $U=5/3$, the asymptotic value for $R_\xi\to 0$.  }
\label{u-rxi-nf2nc4}
\end{figure}

The behavior of the data for different values of the inverse gauge
coupling $\gamma$ shows that the FSS curve is independent of $\gamma$,
at least in a wide interval around $\gamma=0$, as can be seen from
Fig.~\ref{u-rxi-nf2nc3-gamma}, where data for $\gamma=\pm 3$ are
reported. Analogous results are obtained for $N_f=2$ and $N_c=4$, see
Fig.~\ref{u-rxi-nf2nc4}.

\begin{figure}[tbp]
\includegraphics*[scale=\graphicscale]{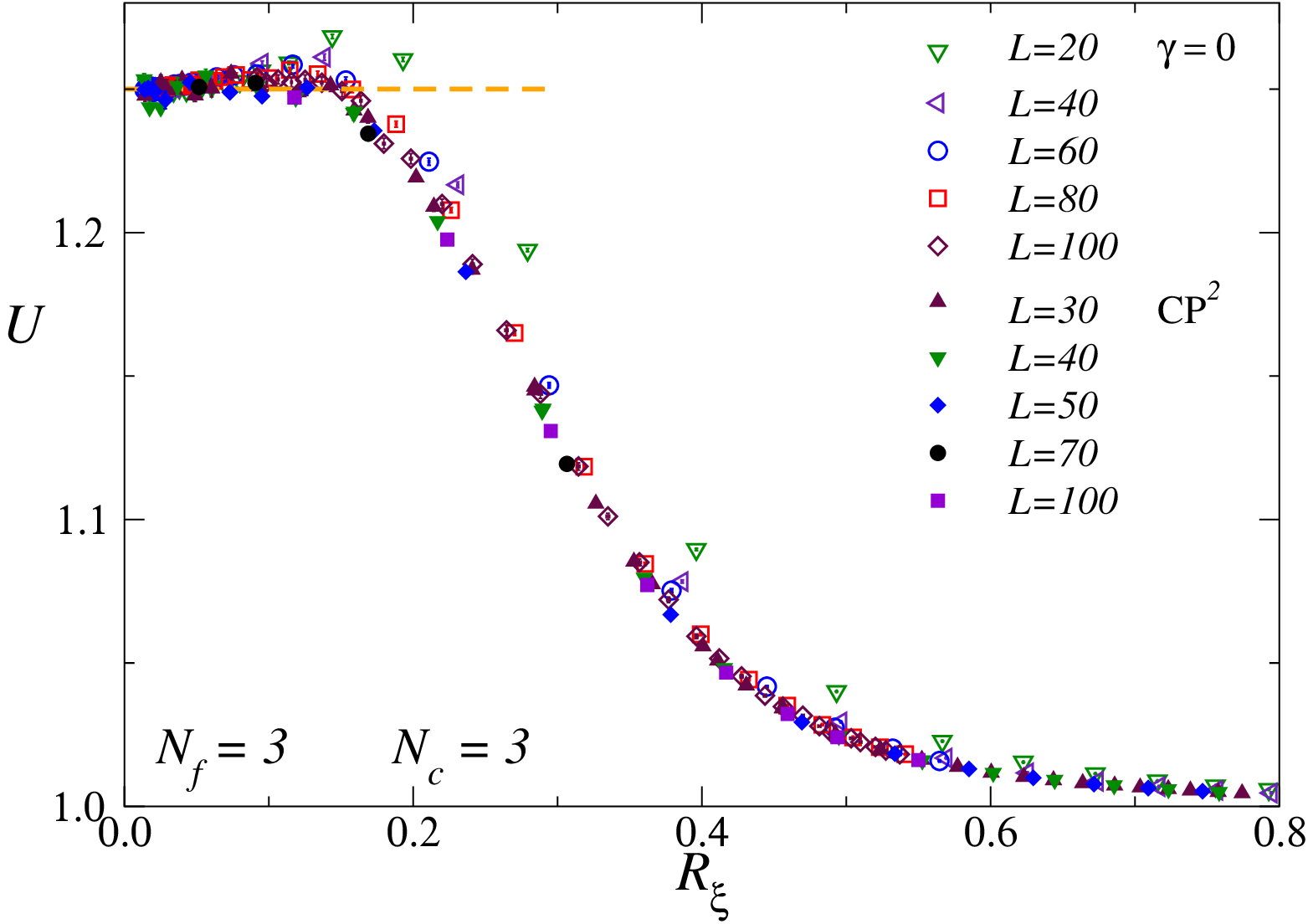}
\includegraphics*[scale=\graphicscale]{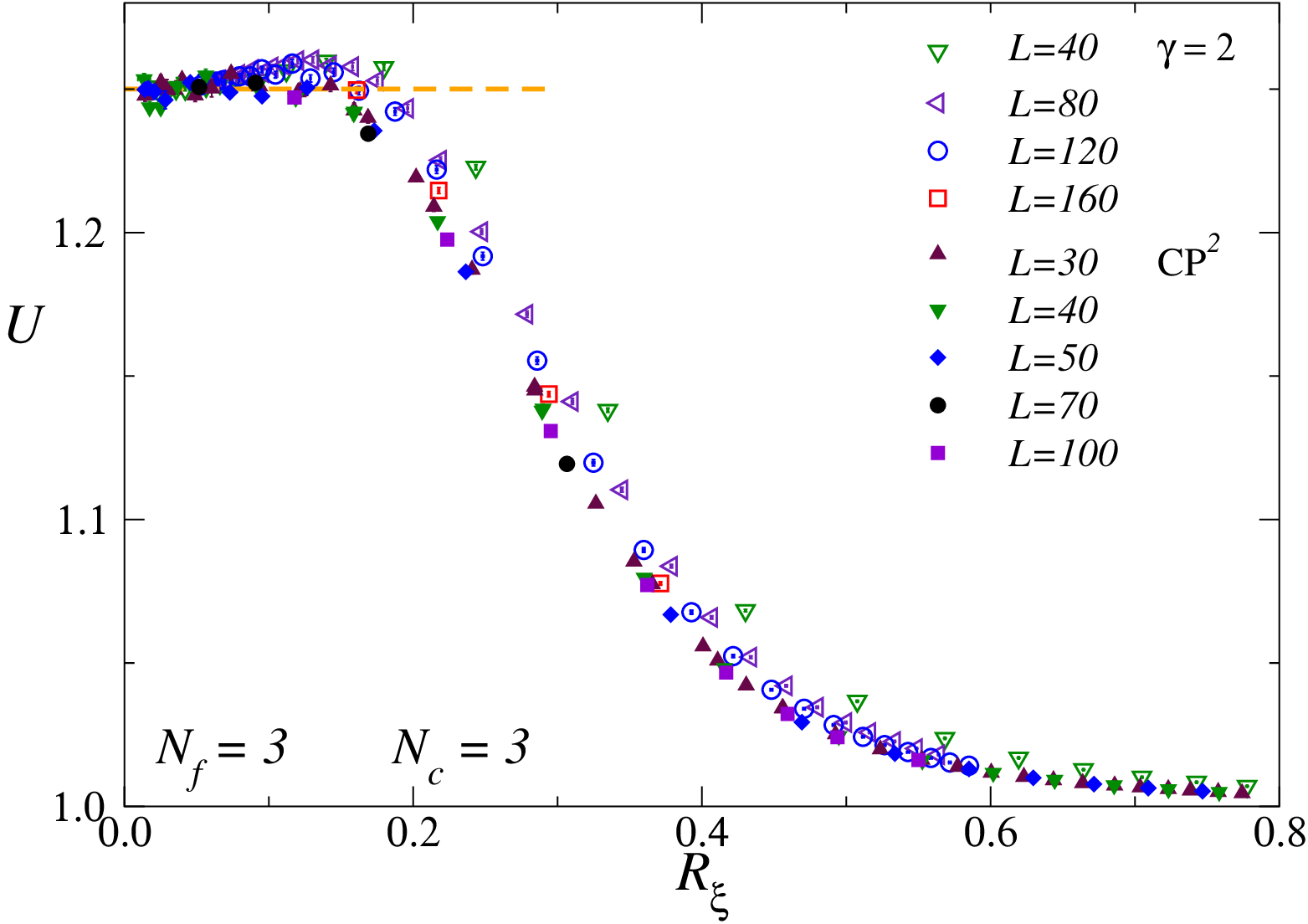}
\caption{Plot of $U$ versus $R_\xi$ for $N_f=3$, $N_c=3$. Results for
  $\gamma=0$ (top) and $\gamma = 2$ (bottom).  Data (empty symbols)
  approach the universal FSS curve of the 2D CP$^2$ universality class
  (CP$^2$ results, taken from Ref.~\cite{BPV-19-2}, are reported with
  full symbols).  The horizontal dashed line corresponds to $U=5/4$,
  the asymptotic value for $R_\xi\to 0$.  }
\label{u-rxi-nf3nc3}
\end{figure}

\begin{figure}[tbp]
\includegraphics*[scale=\graphicscale]{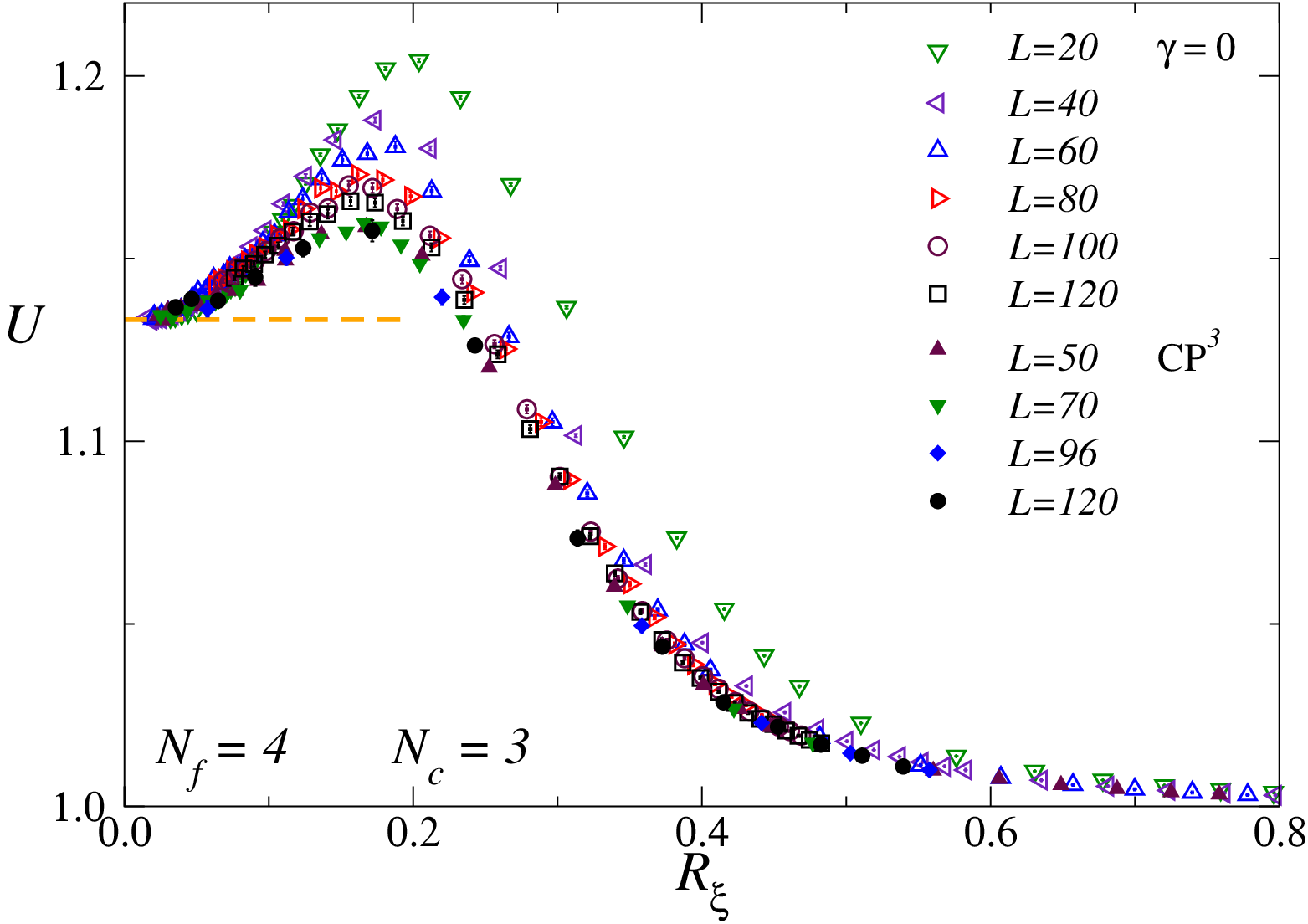}
\caption{Plot of $U$ versus $R_\xi$ for $N_f=4$, $N_c=3$, and
  $\gamma=0$.  Data (empty symbols) Data (empty symbols) approach the
  universal FSS curve of the 2D CP$^3$ universality class (CP$^2$
  results, taken from Ref.~\cite{BPV-19-2}, are reported with full
  symbols).  The horizontal dashed line corresponds to $U=17/15$, the
  asymptotic value for $R_\xi\to 0$. }
\label{u-rxi-nf4nc3}
\end{figure}

These results should be considered as a robust evidence that the
asymptotic low-temperature behavior of two-flavor 
chromodynamics with SU(3) and SU(4) gauge symmetry belongs to the
universality class of the 2D CP$^1$ [equivalently, O(3)] field theory.

In Fig~\ref{u-rxi-nf3nc3} we report results for the three-flavor
lattice theory with SU(3) gauge symmetry. In this case, for both
$\gamma=0$ and $\gamma=2$, data appear to approach the FSS curve of
the 2D CP$^2$ model, obtained in Ref.~\cite{BPV-19-2} by numerical
simulations. This excellent agreement provides a robust indication
that the three-flavor lattice theory with SU(3) gauge theory has the
same asymptotic critical behavior as the 2D CP$^2$ model.  Analogous
results are obtained for $N_f=4$, see Fig.~\ref{u-rxi-nf4nc3} for
results for $\gamma=0$.  The FSS curve appears to approach that of the
2D CP${^3}$ model.  We note that for $N_f=4$ larger scaling
corrections are present.  However, they appear to be consistent with
an $O(L^{-2})$ behavior.

\begin{figure}[tbp]
\includegraphics*[scale=\graphicscale]{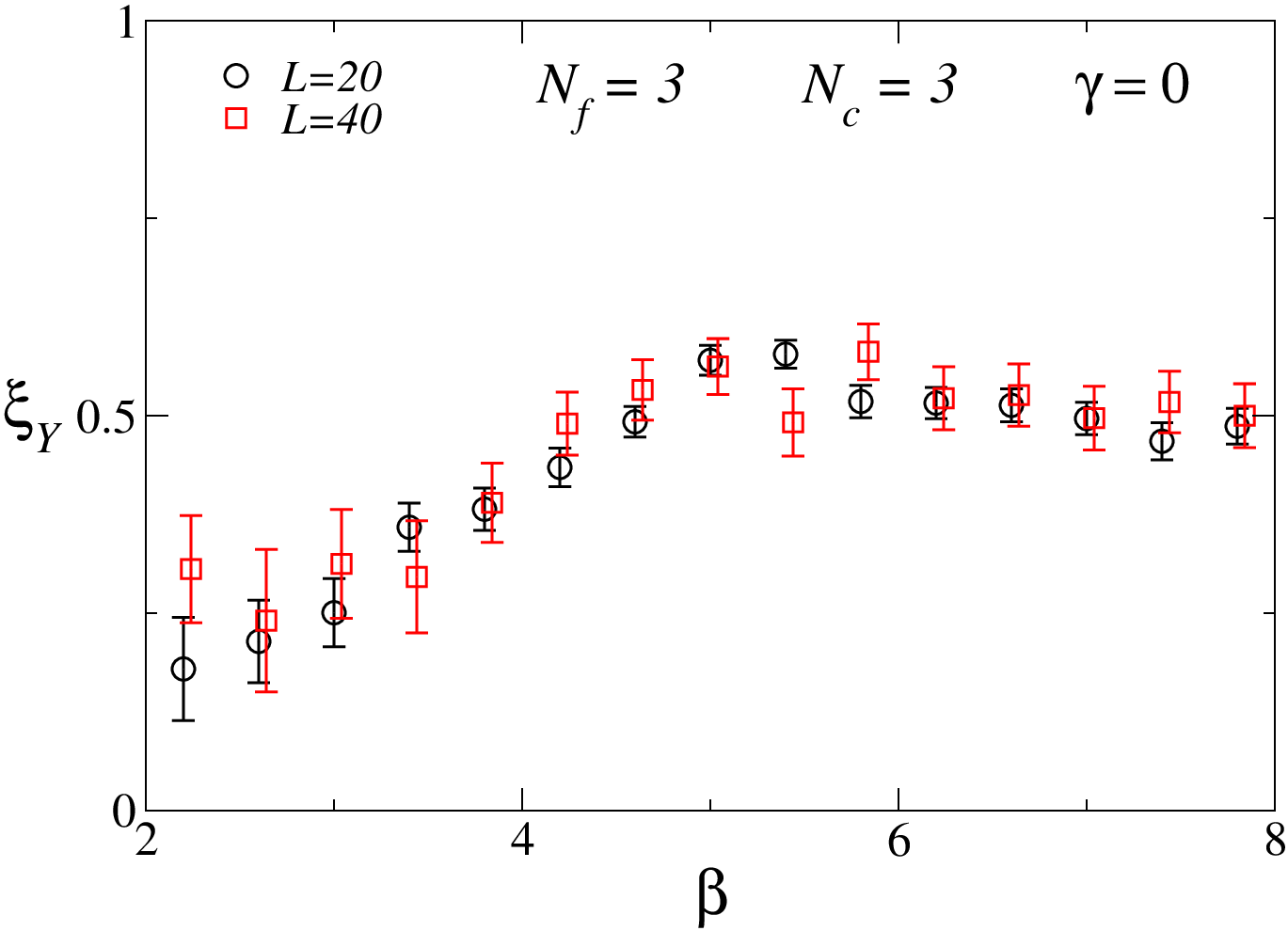}
   \caption{Plot of the correlation length $\xi_Y$ associated with the
     correlation function $\langle \bar{Y}_{\bm x} Y_{\bm y}\rangle$
     versus $\beta$.  The quantity $Y_{\bm x}$ is defined in
     Eq.~\eqref{detof}.  Results for $N_f=N_c=3$ and $\gamma=0$.  }
\label{u1-xi-nf3nc3}
\end{figure}

Up to now we have discussed the critical behavior of $Q$ correlations.
However, note that the model has the additional U(1) global
invariance, Eq.~(\ref{u1varphi}). As we have already discussed, for
$N_f < N_c$ such an invariance is only apparent, but in principle it
may be relevant for $N_f \ge N_c$. To understand its role, we have
studied the behavior of an appropriate order parameter.  As discussed
in Ref.~\cite{BPV-20}, for $N_f=N_c$ an order parameter is provided by
the composite operator
\begin{equation}
Y_{\bm x} = {\rm det} \,\varphi_{\bm x}\,,
\label{detof}
\end{equation}
which is invariant under both the SU($N_c$) gauge transformations
(\ref{gautra}) and the global transformations $\varphi_{\bm x} \to
\varphi_{\bm x} \, V$ with $V\in {\rm SU}(N_f)$. Starting from $Y_{\bm
  x}$, one can define a correlation function
\begin{equation}
G_Y({\bm x} - {\bm y}) = \langle \bar{Y}_{\bm x} {Y}_{\bm y} \rangle,
\end{equation}
and a correlation length $\xi_Y$, using Eq.~(\ref{xidefpb}).  Results
for $\xi_Y$ for $N_f=N_c=3$ and $\gamma=0$ are presented in
Fig.~\ref{u1-xi-nf3nc3}. Apparently, $\xi_Y$ remain finite and very
small ($\xi_Y \approx 0.5$) as $\beta$ increases. The U(1) flavor
modes are clearly not relevant for the critical behavior, which is
completely controlled by the U(1)-invariant modes encoded in $Q_{\bm
  x}$.  The possibility of a U(1) critical behavior, which would imply
the presence of a finite-temperature Berezinskii-Kosterlitz-Thouless
transition~\cite{KT-73,Berezinskii-70,Kosterlitz-74,JKKN-77}, is
excluded by the MC data. The behavior we observe is completely
analogous to what occurs in three dimensions at finite
temperature~\cite{BPV-20}.

\subsection{Universality class of the asymptotic
low-temperature behavior}
\label{univclassesncl2}

The numerical FSS analyses reported above suggest that, for $N_c \ge
3$, the low-temperature asymptotic behavior of scalar chromodynamics
with $N_f$ flavors depends only on $N_f$. Irrespective of the values
of $N_c$ and of $\gamma$, the critical behavior is the same as that of
the 2D CP$^{N_f-1}$ model.

Before presenting further arguments to support such a conclusion, we
recall some features of the 2D CP$^{N-1}$
model~\cite{Witten-79,ZJ-book}.  This is a 2D quantum field theory
defined on a complex projective space, isomorphic to the symmetric
space U($N$)$/[\mathrm{U(1)}\times \mathrm{U}(N-1)]$. Its Lagrangian
reads
\begin{eqnarray}
&&{\cal L} = {1\over 2 g}
 \overline{D_{\mu}{\bm z}}\cdot D_\mu {\bm z}\,, \qquad  
 \bar{\bm z}\cdot {\bm z}=1\,,\qquad
\label{contham}\\  
 &&
 D_\mu =
  \partial_\mu + i A_\mu\,, \qquad
A_\mu = i\bar{{\bm z}}\cdot \partial_\mu {\bm z}\,,
\nonumber
\end{eqnarray}
where ${\bm z}$ is an $N$-component complex field, and $A_\mu$ is a
composite gauge field.  The Lagrangian in invariant under the local
U(1) gauge transformations ${\bm z}({\bm x})\to e^{i\theta({\bm x})}
{\bm z}({\bm x})$, and the global transformations ${\bm z}({\bm x})\to
W {\bm z}({\bm x})$ with $W\in {\rm SU}(N)$. The global invariance
group is SU($N$)$/\mathbb{Z}_N$ (again global transformations
differing by a $\mathbb{Z}_N$ factor are gauge equivalent).

For $N=2$ the CP$^1$ field theory is locally isomorphic to the O(3)
non-linear $\sigma$ model with the identification of the
three-component real vector $s_{\bm x}^a=\sum_{ij} \bar{z}_{\bm x}^i
\sigma_{ij}^{a} z_{\bm x}^j$, where $a=1,2,3$ and $\sigma^{a}$ are the
Pauli matrices.  Various lattice formulations of CP$^{N-1}$ models
have been considered, see, e.g., Refs.~\cite{CR-93,CRV-92}. The
simplest formulation is
\begin{equation}
S_{CP} = - J \sum_{{\bm x}\,\mu} | \bar{\bm{z}}_{\bm x} \cdot {\bm
  z}_{{\bm x}+\hat\mu} |^2 = - J \sum_{{\bm x}\mu} {\rm Tr} \,{\cal
  P}_{\bm x} {\cal P}_{{\bm x}+\hat{\mu}} \,,
\label{hcpn}
\end{equation}
where
\begin{equation}
{\cal P}_{\bm x}^{ab} = \bar{z}_{\bm x}^a z_{\bm x}^b\,
\label{pxdef}
\end{equation}
is a projector, i.e., it satisfies ${\cal P}_{\bm x} = {\cal P}_{\bm
  x}^2$.  This explicitly shows that 2D CP$^{N-1}$ theories describe
the dynamics of projectors on $N$-dimensional complex spaces.
CP$^{N-1}$ models can also be obtained by considering the action
(\ref{hgauge}) with $\gamma = 0$, $z$ replacing the field $\varphi$
and using U(1) gauge fields $U_{{\bm x},\mu}$.

In 2D CP$^{N-1}$ models correlations are always short-ranged at finite
$\beta$ \cite{MW-66}. A critical behavior is only observed for $T\to
0$. In this limit the correlation length increases exponentially
as~\cite{ZJ-book,CR-93}
\begin{equation}
\xi \sim T^p e^{c/T}\,.
\label{xibeta}
\end{equation}
This behavior is related to the asymptotic-freedom property of these
models, which is shared with quantum chromodynamics, the four
dimensional theory of strong interactions.  An analogous exponential
behavior is expected to characterize all statistical lattice theories
belonging to the same universality class, therefore also the 2D
$N$-component Abelian-Higgs lattice model~\cite{BPV-19-2} (which is a
lattice version of scalar electrodynamics), and, as we shall argue,
the 2D scalar chromodynamics with $N_f$ flavors and $N_c\ge 3$.

The numerical results reported in Sec.~\ref{resnf2ncl2} show that the
asymptotic low-temperature behavior of the model with $N_c\ge 3$ is
the same as that of the 2D CP$^{N_f-1}$ models or, equivalently, of
the 2D $N_f$-component Abelian-Higgs model with a U(1) gauge
symmetry. This is certainly quite surprising.  We shall now argue that
the correspondence is strictly related to the identical nature of the
minimum-energy configurations, which represent the background for the
spin waves that are responsible for the zero-temperature critical
behavior.

The nature of the minimum-energy configurations is discussed in
App.~\ref{largebeta}.  For $\gamma \ge 0$, such configurations are
those for which
\begin{equation}
{\rm Re}\,{\rm Tr} \,\varphi_{\bm x}^\dagger  
U_{{\bm x},\mu} \varphi_{{\bm x}+\hat\mu} = 1\,
\label{minene}
\end{equation}
on each lattice link. In the appendix, by combining exact and
numerical results, we show that, for $\beta\to\infty$ and $N_c\ge 3$,
by appropriately fixing the gauge, the configurations that dominate
the statistical average have the form
\begin{equation}
\Pi_{\bm x} = 
\begin{pmatrix} V & 0 \\
                0 & 1 
\end{pmatrix}
\label{Pi-largebeta-testo}
\end{equation}
where $V$ is an SU($N_c-1$) matrix, and 
\begin{equation}
\begin{array}{ll}
 \varphi^{af} = 0 \qquad & {a < N_c}\, ,  \\
 \varphi^{af} = z^f \qquad & {a = N_c}\, ,
\end{array}
\label{phi-largebeta-testo}
\end{equation}
where $z^f$ is a unit-length $N_f$-dimensional vector. In other words,
the analysis shows that gauge and $\varphi$ fields completely
decouple. Moreover, the $\varphi$ field becomes equivalent to a single
unit-length $N_f$-dimensional vector, which is the fundamental field
of the CP$^{N_f-1}$ model.  Stated differently, the operator $P_{\bm
  x}$ becomes a projector, i.e., satisfies $P_{\bm x}^2 = P_{\bm x}$,
for $T\to 0$.  However, we cannot yet, at this point, argue that the
large-$\beta$ behavior of scalar chromodynamics and of the
CP$^{N_f-1}$ model is the same, because in our factorization there is
no U(1) gauge symmetry. However, our numerical data also show that the
critical behavior is only associated with the order parameter $Q_{\bm
  x}$: the U(1) modes do not order in the large-$\beta$ limit. This is
also confirmed by the detailed analysis of the low-temperature
configurations presented in Ref.~\cite{BPV-20}. Therefore, in the
effective theory we can quotient out the U(1) degrees of freedom,
which are irrelevant for the behavior of the order parameter $Q_{\bm
  x}$, i.e., we can reintroduce the U(1) gauge symmetry. If this
occurs, scalar chromodynamics and CP$^{N_f-1}$ model are expected to
have the same critical large-$\beta$ behavior.

It is interesting to observe that CP$^{N_f-1}$ behavior has also been
observed for several negative values of $\gamma$. This is not an
obvious result, as the system is frustrated. Also in this case, the
result is explained by the nature of the low-temperature
configurations. As discussed in App.~\ref{largebeta} for a specific
value of $\gamma$, $\gamma=-1$, the relevant configurations can again
be parametrized as in Eq.~(\ref{phi-largebeta-testo}), modulo gauge
transformations.

This phenomenological argument explains the numerical evidence that
the asymptotic zero-temperature behaviors for $N_c\ge 3$ is the same
as that of the CP$^{N_f-1}$ continuum theory.  Note that this scenario
does not apply to nonabelian gauge theories with $N_c=2$. As discussed
in App.~\ref{largebeta}, the typical low-temperature configurations
cannot be parametrized as in Eq.~(\ref{phi-largebeta-testo}), implying
a different critical behavior.  We shall argue that it corresponds to
that of the 2D Sp($N_f)$ field theories.

\section{SU(2) gauge models}
\label{resnc2}

We now discuss the behavior of models with SU(2) gauge symmetry.  In
this case the global symmetry group ~\cite{BPV-19,BPV-20} is
Sp($N_f$)/$\mathbb{Z}_2$. In the two-flavor case, because of the
isomorphism Sp(2)$/\mathbb{Z}_2=$SO(5), an O(5) symmetry emerges.
Because of the symmetry enlargement, the order-parameter field is the
$2N_f\times 2N_f$ matrix
\begin{equation}
{\cal T}_{\bm x}^{lm} = 
\sum_a \overline{\Gamma}_{\bm x}^{al} \Gamma_{\bm x}^{am} -
     {\delta^{lm} \over N_f} \,,
\label{Tdef}
\end{equation}
where the matrix $\Gamma_{\bm x}$ is defined in Eq.~(\ref{Gammadef}).
If $f,g=1,...,N_f$, ${\cal T}_{\bm x}^{lm}$ can be written in the
block form
\begin{equation} \label{tdrel}
\begin{aligned}
&{\cal T}_{\bm x}^{f,g}=Q_{\bm x}^{fg}\,
& &{\cal T}_{\bm x}^{f,g+N_f} = \bar{D}^{fg}\, \\
&{\cal T}_{\bm x}^{f+N_f,g} = -D^{fg}\,
& &{\cal T}_{\bm x}^{f+N_f,g+N_f}=Q_{\bm x}^{gf}\,
\end{aligned}
\end{equation}
where
\begin{equation}
D_{\bm x}^{fg} = \sum_{ab} \epsilon^{ab} \varphi_{\bm x}^{af}
\varphi_{\bm x}^{bg}\, .
\end{equation}
The order parameter ${\cal T}_{\bm x}$ is hermitian and satisfies
\begin{equation}
J \overline{\cal T}_{\bm x} J + {\cal T}_{\bm x} = 0\,,
\label{tcostr}
\end{equation}
where the matrix $J$ is defined in Eq.~(\ref{mscond}). 

For $N_f=2$ the matrix ${\cal T}_{\bm x}$ can be parametrized by a
five-dimensional real vector ${\bm \Phi}_{\bm x}$. The first three
components are given by
\begin{equation}
\Phi_{\bm x}^k \equiv \sum_{fg} \sigma^k_{fg} Q_{\bm x}^{fg}\,,
\quad k=1,2,3\,,
\label{phidef}
\end{equation}
while the fourth and fifth component are the real and imaginary parts
of
\begin{equation}
{1\over2} \sum_{fg}\epsilon_{fg} D^{fg}_{\bm x}
\equiv 
\Phi_{\bm x}^4 + i\Phi_{\bm x}^5 \,.
\label{psidef}
\end{equation}
The parametrization of ${\cal T}_{\bm x}$ in terms of $\Phi_{\bm x}$
effectively implements the isomorphism between the Sp(2)/${\mathbb
  Z}_2$ and the SO(5) groups, since an Sp(2) transformation of ${\cal
  T}_{\bm x}$ maps to an SO(5) rotation of ${\bm\Phi}_{\bm x}$.
Moreover, the unit-length condition for $\varphi$ implies
\begin{equation}
{\bm \Phi}_{\bm x}\cdot {\bm \Phi}_{\bm x} = 1\,.
\label{normphipsi}
\end{equation} 

\begin{figure}[tbp]
\includegraphics*[scale=\graphicscale]{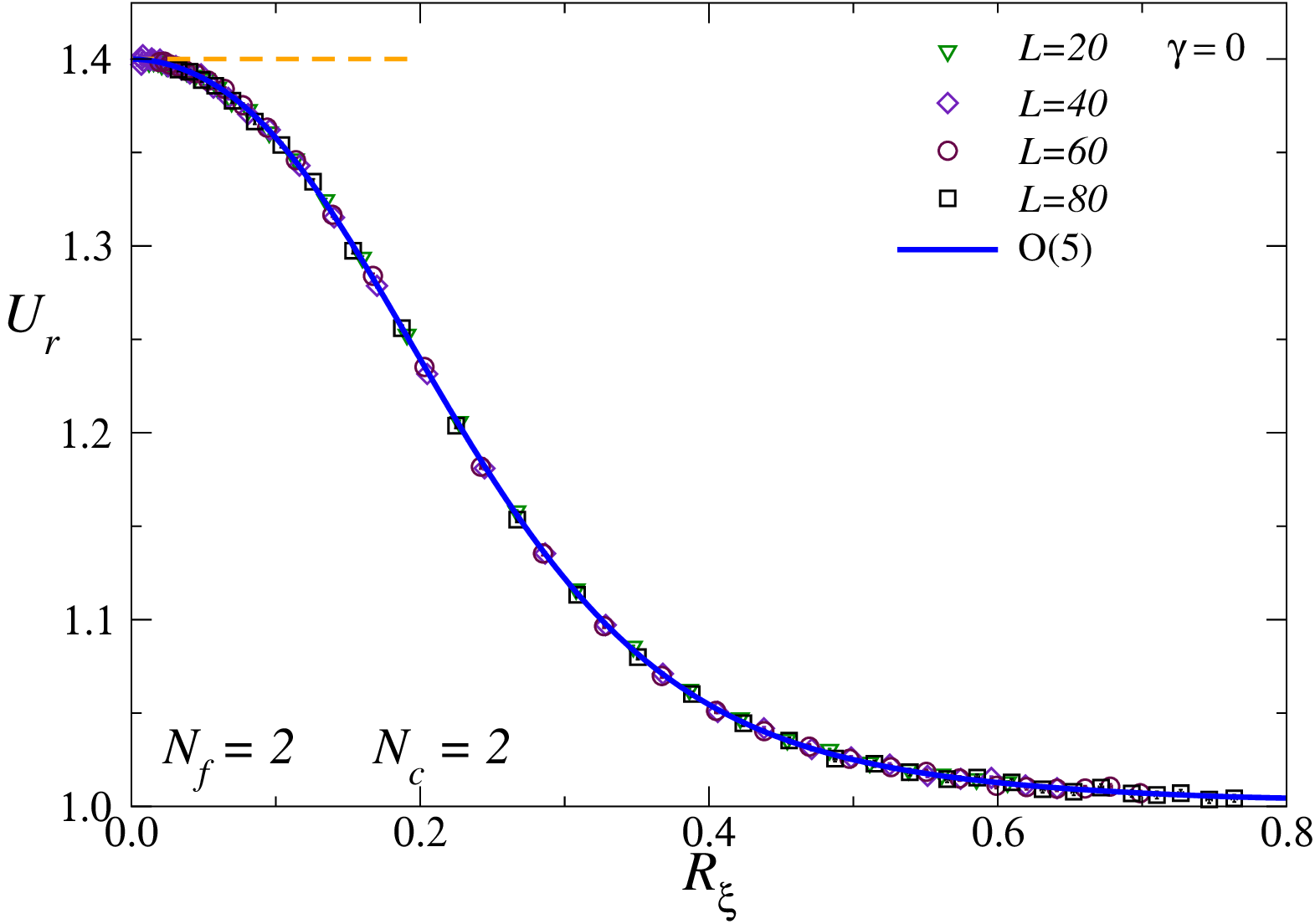}
\caption{Plot of $U_r$ versus $R_\xi$ for $N_f=2$, $N_c=2$, and
  $\gamma=0$.  The data approach a universal FSS curve, which
  corresponds to that of the standard O(5) nearest-neighbor vector
  model (full line, see Fig.~\ref{u-rxi-o5}).  The horizontal dashed
  line corresponds to the asymptotic value $U_r=7/5$ for $R_\xi\to
  0$.}
\label{u-rxi-nf2nc2}
\end{figure}

\begin{figure}[tbp]
\includegraphics*[scale=\graphicscale]{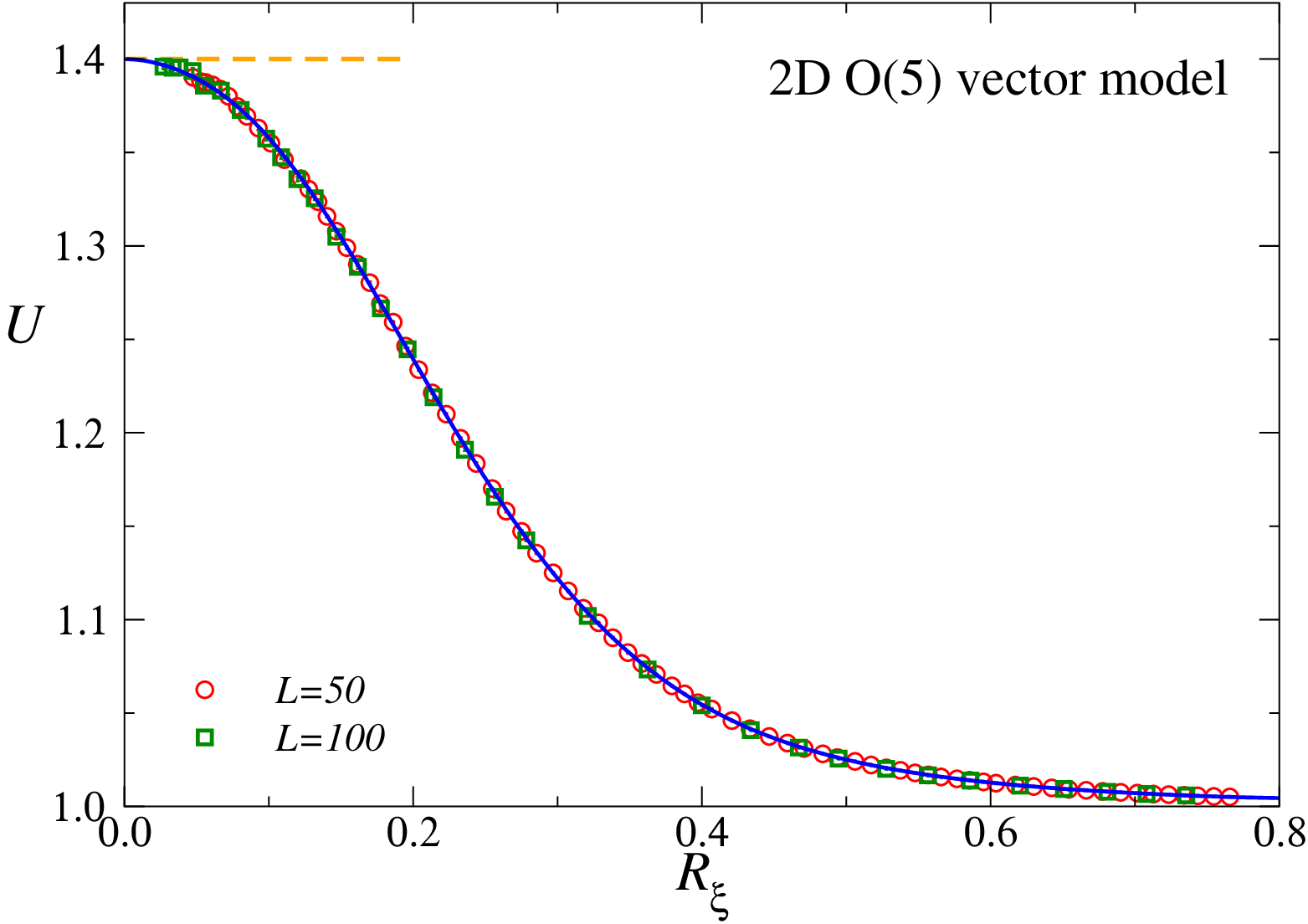}
\caption{Plot of $U$ versus $R_\xi$ for the O(5) vector universality
  class, as obtained by MC simulations of the nearest-neighbor O(5)
  vector lattice model. The full line~\cite{footnote-interpolation},
  is an interpolation the MC data up to $R_\xi\lesssim 0.8$. It
  provides an approximation of the universal FSS curve, with an
  accuracy smaller than 0.5\% (we include the uncertainty arising from
  scaling corrections). The horizontal dashed line corresponds to the
  asymptotic value $U=7/5$ for $R_\xi\to 0$; $U\to 1$ for
  $R_\xi\to\infty$.  }
\label{u-rxi-o5}
\end{figure}

\begin{figure}[tbp]
\includegraphics*[scale=\graphicscale]{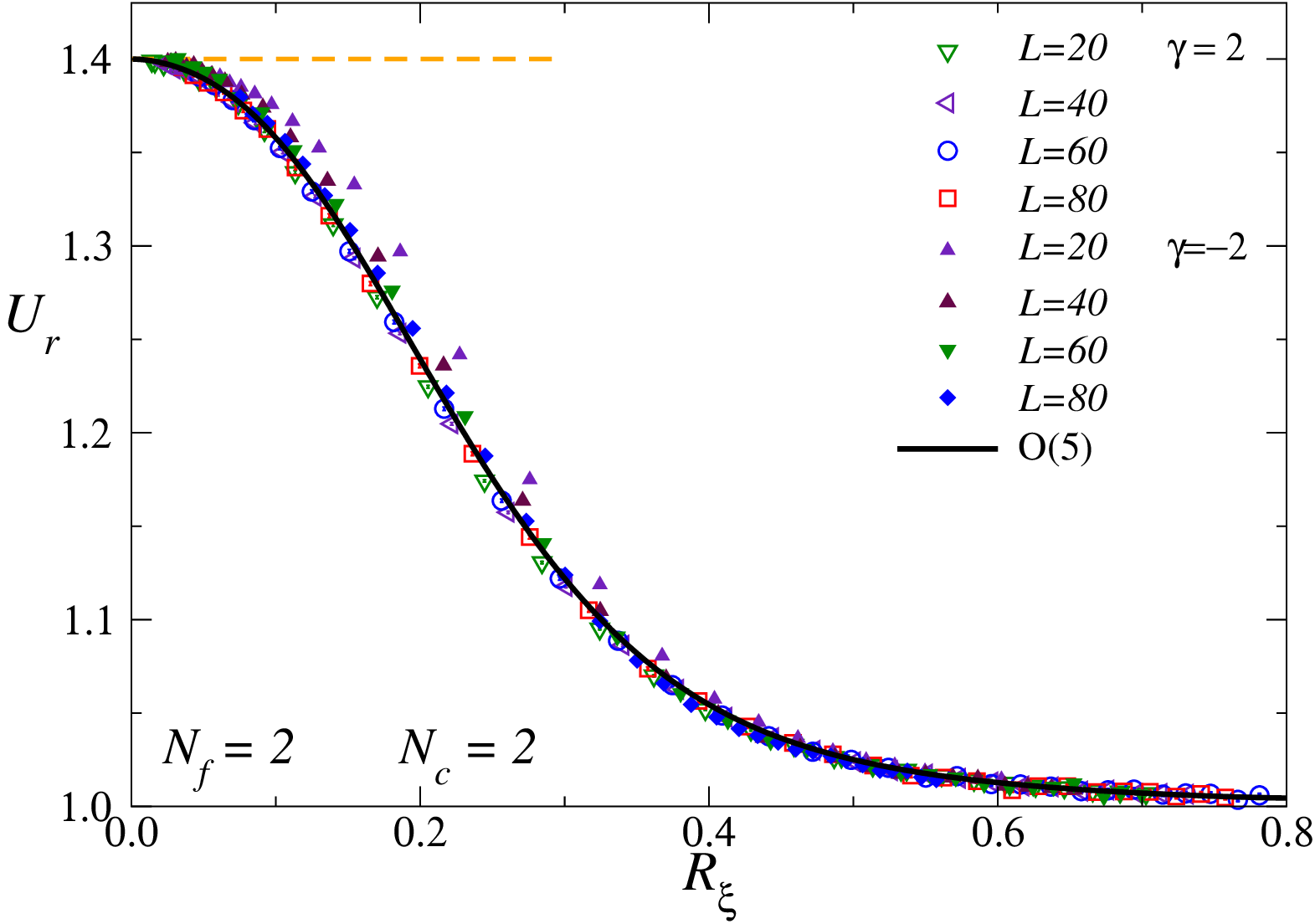}
\caption{Plot of $U_r$ versus $R_\xi$ for $N_f=2$, $N_c=2$, and
  $\gamma=\pm 2$.  The data approach the O(5) FSS curve in the large
  $L$ limit (full line~\cite{footnote-interpolation}).  The horizontal
  dashed line corresponds to the asymptotic value $U_r=7/5$ for
  $R_\xi\to 0$.}
\label{u-rxi-nf2nc2-gamma}
\end{figure}

The discussion of the previous section leads us to conjecture that the
global Sp($N_f$)$/\mathbb{Z}_2$ symmetry uniquely determines the
asymptotic zero-temperature critical behavior.  For $N_f=2$ this would
imply that the SU(2) gauge theory has the same zero-temperature
behavior of the O(5) vector model.  An analogous conjecture proved to
be true in the three-dimensional case~\cite{BPV-19,BPV-20}.  To
perform the correct universality check for $N_f=2$, as discussed in
detail in Ref.~\cite{BPV-20}, it is important to consider a Binder
parameter in the SU(2) gauge theory that maps onto the usual vector
O(5) Binder parameter under the isomorphism Sp($N_2$)$/\mathbb{Z}_2
\to $ SO(5).  The Binder parameter $U$ defined in
Eq.~(\ref{binderdef}) is not the appropriate one since it only
involves three components of $\Phi_{\bm x}^k$, see
Eq.~(\ref{phidef}). A straightforward group-theory computation shows
that the correct correspondence is achieved by defining the related
quantity~\cite{BPV-19,BPV-20}
\begin{equation}
U_r = {21\over 25}\, U\,.
\label{urdef}
\end{equation}
As for $R_\xi$, the quantity computed using Eq.~(\ref{xidefpb}) corresponds 
exactly to the analogous quantity computed in the O(5) vector model.

The results shown in Fig.~\ref{u-rxi-nf2nc2} clearly support the
conjecture. Indeed, the MC data of $U_r$ collapse (without appreciable
scaling violations) on a unique curve when plotted versus $R_\xi$,
which is consistent with that of the Binder parameter $U$ versus
$R_\xi$ for the 2D O(5) vector model (with $U$ and $R_\xi$ defined
analogously in terms of $\Phi$ correlations~\cite{BPV-19-2}).  The
O(5) FSS curve is obtained by MC simulations (using the cluster
algorithm) of the nearest-neighbor O(5) vector model (\ref{ullimits}),
see Fig.~\ref{u-rxi-o5}. Again the role of the inverse gauge coupling
is irrelevant. It does not change the universal features of the
low-temperature asymptotic behavior, as shown in
Fig.~\ref{u-rxi-nf2nc2-gamma}, where we report results for $\gamma=2$
and $-2$.

The numerical analysis reported above leads us to conjecture that the
low-temperature asymptotic behavior of the scalar SU(2) chromodynamics
with $N_f$ flavors belongs to the universality class associated with
the 2D Sp($N_f$) field theory. The fundamental field is a complex
$2N_f\times 2N_f$ order-parameter field $\Psi_{\bm x}$, which formally
represents a coarse-grained versio of ${\cal T}_{\bm x}$, defined in
Eq.~(\ref{Tdef}). It is hermitian, traceless, and satisfies
Eq.~(\ref{tcostr}). If we write
\begin{equation}
\Psi =
\left(\begin{array}{cc} A_1 & A_2 \\ A_3 &
  A_4\end{array}\right)\,,
\label{psia}
\end{equation}
where $A_i$ are $N_f\times N_f$ matrix fields, the conditions required
are that $A_1$ is hermitian and traceless, $A_3$ is antisymmetric,
$A_4 = \bar{A}_1$, and $A_3=-\bar{A}_2$.  The corresponding 2D field
theory is defined by the Lagrangian
\begin{eqnarray}
{\cal L}_{\rm Sp} = {1\over g} \,{\rm Tr}\left[
 \partial_{\mu} \Psi^\dagger\, \partial_{\mu} \Psi\right]\,,\qquad
 {\rm Tr} \,\Psi^\dagger \Psi =  1\,.
\label{conthamsp}
\end{eqnarray}
For $N_f=2$, using the correspondence 
\begin{eqnarray}
&&A_1 ={1\over 2}
\left(\begin{array}{cc} \Phi^3 & \Phi^1-i\Phi^2 \\ \Phi^1+i\Phi^2 &
  -\Phi^3\end{array}\right)\,, \label{a1nf2}\\
&&A_2 = {1\over 2}
\left(\begin{array}{cc} 0 & \Phi^4+i\Phi^5 \\ -\Phi^4-i\Phi^5 &
  0\end{array}\right)\,, \nonumber
\end{eqnarray}
one can easily show that the Sp(2) field theory is  equivalent 
to the O(5) $\sigma$-model with Lagrangian
\begin{eqnarray}
{\cal L}_{\rm O} = {1\over  g} \,
 \partial_{\mu} \Phi \cdot \partial_{\mu} \Phi \,,\qquad
 \Phi \cdot \Phi =  1\,.
\label{conthamspOM}
\end{eqnarray}

\section{Conclusions}
\label{conclu}

We have studied a 2D lattice nonabelian gauge model with
multicomponent scalar fields, focusing on the role that global and
local nonabelian gauge symmetries play in determining the universal
features of the asymptotic low-temperature behavior.  The lattice
model we consider is obtained by partially gauging a maximally
O($M$)-symmetric multicomponent scalar model, using the Wilson lattice
approach. The resulting theory is locally invariant under SU($N_c$)
gauge transformations ($N_c$ is the number of colors) and globally
invariant under SU($N_f$) transformations ($N_f$ is the number of
flavors).  The fields belong to the coset $S^M$/SU($N_c$), where $M=2
N_f N_c$ and $S^M$ is the $M$-dimensional sphere. The model is always
disordered at finite temperature, in agreement with the Mermin-Wagner
theorem \cite{MW-66}.  However, it develops a critical behavior in the
zero-temperature limit.  The corresponding universal features are
determined by means of numerical analyses of the FSS behavior in the
zero-temperature limit.

We observe universality with respect to the inverse gauge coupling
$\gamma$ that parametrizes the strength of the gauge kinetic term, see
Eq.~(\ref{hgauge}). The RG flow is always controlled by the infinite
gauge-coupling fixed point, corresponding to $\gamma=0$, as it also
occurs in three dimensions~\cite{BPV-19,BPV-20}, and in 2D and 3D
models characterized by an abelian U(1) gauge
symmetry~\cite{BPV-19-2,PV-19}. Indeed, models corresponding to
different values of $\gamma$ have the same universal behavior for
$T\to 0$, at least in a large interval around $\gamma=0$.  We
conjecture that the same critical behavior is obtained for all
positive finite values of $\gamma$, since, by increasing $\gamma$, we
do not expect any qualitative change in the structure of the
minimum-energy configurations that control the statistical average.
On the other hand, the behavior for negative values of $\gamma$, i.e.,
when the system is frustrated, is not completely understood.
Therefore, we cannot exclude that the behavior changes for large
negative values of $\gamma$. This issue remains an open problem. It is
important to note that by considering a positive value of $\gamma$, we
are effectively investigating the behavior close to the multicritical
point $\beta = \infty$ and $\beta_g = \beta\gamma = \infty$. Our
results show that approching the point along the lines $\beta_g/\beta
= \gamma$ does not change the universal features of the asymptotic
behavior. However, we expect that, by increasing $\beta_g$ faster than
$\beta$ (in a well-specified way), one can observe a radical change in
the critical behavior. For instance, if we take first the limit
$\beta_g\to\infty$ at fixed finite $\beta$ and then the limit
$\beta\to \infty$, the model becomes equivalent to the standard O($M$)
vector model, characterized by a different asymptotic low-temperature
behavior.

The numerical results and theoretical arguments presented in this
paper suggest the existence of a wide universality class
characterizing 2D lattice abelian and nonabelian gauge models, which
only depends on the global symmetry of the model.  The gauge group
does not apparently play any particular role.  Indeed, we report
numerical evidence that, for any $N_c$, the asymptotic low-temperature
behavior of the multiflavor scalar gauge theory (\ref{hgauge}) belongs
to the universality class of the 2D CP$^{N_f-1}$ model.  This also
implies that it has the same universal features of the $N_f$-component
lattice scalar electrodynamics (abelian Higgs model)
\cite{BPV-19-2}. It is important to note that the global symmetry
group of model (\ref{hgauge}) is U($N_f$), while the global symmetry
group of the CP$^{N_f-1}$ model is SU($N_f$) (we disregard here
discrete subgroups), so that the global symmetry group of the two
models differs by a U(1) flavor group. As we have discussed in
Ref.~\cite{BPV-20}, the U(1) symmetry is only apparent for $N_f <
N_c$.  and therefore, the symmetry groups of scalar chromodynamics and
of the CP$^{N_f-1}$ model are the same for $N_f < N_c$.  This U(1) symmetry
is instead present for $N_f \ge N_c$.  However, our numerical results
indicate that the U(1) flavor symmetry does not play any role in model
(\ref{hgauge}). The universal critical behavior is only associated
with the U(1)-invariant modes that are encoded in the local bilinear
operator $Q_{\bm x}$, so that the global symmetry group that
determines the asymptotic behavior is SU($N_f$).  Note, however, that
the decoupling of the U(1) flavor modes may not be true in other
models with the same global and local symmetries. If the U(1) modes
become critical, a different critical behavior might be observed. This
issue deserves further investigations.

For $N_c = 2$, the global symmetry group changes: The action is
invariant under Sp($N_f$) transformations.  In this case the
asymptotic low-temperature behavior is expected to be described by the
Sp($N_f)$ continuum theory.  We have numerically checked it for the
two-flavor model, for which the global symmetry Sp(2)$/{\mathbb
  Z}_2\simeq$ SO(5).

Our results lead us to conjecture that the RG flow of the 2D
multiflavor lattice scalar chromodynamics in which the fields belong
to the coset $S^M$/SU($N_c$), where $M=2N_cN_f$, is asymptotically
controlled by the 2D statistical field theories associated with the
symmetric spaces~\cite{BHZ-80,ZJ-book} that are invariant under
SU($N_f$) (for $N_c\ge 3$) or Sp($N_f$) (for $N_c=2$) global
transformations. These symmetry groups are the same invariance groups
of scalar chromodynamics, apart from a U(1) flavor symmetry that is
present for $N_f \ge N_c > 2$, which does not play any role in
determining the asymptotic behavior of the model.

This conjecture may be further extended to models with different
global and local symmetry groups, for instance, to those considered in
Refs.~\cite{BHZ-80,PRV-01}. It would be interesting to verify whether
generic nonabelian models have an asymptotic critical behavior which
is the same as that of the model defined on a symmetric space that has
the same global symmetry group.  This issue deserves further
investigations.

\bigskip

\emph{Acknowledgement}.
Numerical simulations have been performed on the CSN4 cluster of the
Scientific Computing Center at INFN-PISA.

\appendix

\section{Minimum-energy configurations}
\label{largebeta}

In this appendix we identify the minimum-energy configurations for the
action (\ref{hgauge}), summarizing the main arguments reported in
Ref.~\cite{BPV-20} and extending them to the $\gamma\neq0$ case.

\subsection{Behavior for  $\gamma=0$}

We start by considering the simplest case $\gamma=0$. The global
minimum is obtained by the configurations that satisfy the maximum
condition
\begin{eqnarray}
{\rm Re}\,{\rm Tr} \,\varphi_{\bm x}^\dagger  
U_{{\bm x},\mu} \varphi_{{\bm x}+\hat\mu} = 1\,
\label{minenea}
\end{eqnarray}
on each link. This condition is trivially satisfied if
\begin{equation}
\varphi_{{\bm x}+\hat{\mu}} = U_{{\bm x},\mu}^\dagger \varphi_{\bm x}\,,
\label{mincond}
\end{equation}
which implies $Q_{\bm x} = Q_{{\bm x}+\hat{\mu}}$, 
and, thus, the breaking of the SU($N_f$)$/\mathbb{Z}_{N_c}$ symmetry for
$\beta\to\infty$.

If we apply repeatedly the relation (\ref{mincond}) along a plaquette,
we obtain the consistency condition
\begin{equation}
\varphi_{\bm x} = \Pi_{\bm x} \,\varphi_{\bm x}, 
\label{mincond2}
\end{equation}
where $\Pi_{\bm x}$ is the plaquette operator (\ref{plaquette}). 

For $N_c = 2$, Eq.~(\ref{mincond2}) implies that $\Pi_{\bm x}$ is the
identity matrix. The same argument allows us to show that also all
Polyakov loops can be reduced to the identity matrix, so that all
gauge configurations are trivial.  The energy-minimum configurations
can therefore be written as
\begin{eqnarray}
\varphi_{\bm x}^{af} = W_{\bm x}^{ab} A^{bf}\,,\qquad
U_{{\bm x},\mu} = W_{\bm x} W_{{\bm x}+\hat\mu}^\dagger\,,
\label{minconf}
\end{eqnarray}
where $A^{af}$ is a generic space-independent $2\times N_f$ complex
matrix satisfying ${\rm Tr} A^\dagger A=1$, and $W_{\bm x}\in {\rm
  SU}(2)$.  To verify these conclusions we have performed simulations
on a $4^2$ lattice for very large values of $\beta$ ($\beta$ varies
between 30 and 100).  The results are reported in
Table~\ref{tablargebeta}. As expected
\begin{equation}
 {\rm Re} \, {\rm Tr}\, \Pi_{{\bm x}} = N_c.
\label{plaqmincond}
\end{equation}
We have also computed $U$ and $\langle \hbox{Tr }P^2_{\bm x}\rangle$,
confirming the predictions of Ref.~\cite{BPV-20} obtained only
assuming that the configurations of minimal energy are those of the
form (\ref{minconf}):
\begin{equation}
\langle {\rm Tr}\, P_{\bm x} ^2 \rangle \equiv
{N_f + N_c\over 1 + N_f N_c} \,,
\label{myguess}
\end{equation} 
and
\begin{equation}
U
 =\frac{(1+N_f N_c) (N_f N_c + 4 N_f^2 + N_f^3 N_c - 6)}{
   (N_f^2-1) (2+N_f N_c) (3+N_f N_c)},
\label{LTvalue}
\end{equation}
with $N_c = 2$. 

\begin{table*}
\begin{tabular}{cllllll}
\hline\hline
$(N_c, N_f)$ & 
    \multicolumn{1}{c}{$\langle \mathrm{Re\,Tr}\,\Pi_{\bm x}\rangle/N_c$}   & 
    \multicolumn{1}{c}{$S_g/(2N_f)$}  &   \multicolumn{1}{c}{$U$}   &
Eq.~(\ref{LTvalue})   & $\langle \mathrm{Tr }\, P^2_{\bm x}\rangle$ & 
  Eq.~\eqref{myguess}   \\ \hline
(2, 2)       & 1.00004(2)  & $-$1.00000(1)   & 1.191(2)     & 1.19048$\ldots$ & 0.800(1)   & 0.8                  \\ 
(2, 3)       & 1.00000(2)  & $-$0.999994(7)  & 1.094(1)     & 1.09375$\ldots$ & 0.714(1)   & 0.714286$\ldots$     \\ \hline
(3, 2)       & 0.3421(3)   & $-$0.999998(8)  & 1.0000000(5) &                 & 1.00002(2) &                  \\
(3, 3)       & 0.3501(4)   & $-$1.00000(1)   & 1.000000(2)  &                 & 0.99999(2) &                  \\ \hline
(4, 2)       & 0.2534(2)   & $-$0.999997(7)  & 1.0000000(5) &                 & 1.00001(2) &                  \\
\hline\hline
\end{tabular}
\caption{Asymptotic values for $\beta\to\infty$ on a $4^2$ lattice for 
$\gamma=0$.}
\label{tablargebeta}
\end{table*}

For $N_c\ge 3$, the minimum-energy condition
(\ref{mincond2}) has several classes of different solutions. 
If $\Pi_{\bm x}$ satisfies Eq~(\ref{mincond2}),
we can always write it as 
\begin{equation}
\Pi_{\bm x} = V\oplus 1 = 
\begin{pmatrix} V & 0 \\
                0 & 1 
\end{pmatrix}
\label{Pi-largebeta}
\end{equation}
where $V$ is an SU($N_c-1$) matrix, modulo a gauge transformation. The
corresponding configurations of the fields $\varphi_x$ depend on the
structure of the matrix $V$. If $V$ is a generic unitary matrix which
does not have unit eigenvalues, Eq.~(\ref{mincond2}) implies that the
field $\varphi$ is necessarily given by
\begin{equation}
\begin{array}{ll}
 \varphi^{af} = 0 \qquad & {a < N_c}\, ,  \\
 \varphi^{af} = z^f \qquad & {a = N_c}\, ,
\end{array}
\label{phi-largebeta}
\end{equation}
where $z^f$ is a unit $N_f$-dimensional vector.  Different $\varphi$
configurations are only possible if $V$ has some unit eigenvalues. For
instance, if $V = V_1 \oplus 1$, with $V_1$ belonging to the
SU($N_c-2$) subroup, then the $\varphi$ field configurations of the
form
\begin{equation}
\begin{array}{ll}
 \varphi^{af} = 0 \qquad & {a < N_c-1}\, , \\ \varphi^{af} = w^f
 \qquad & {a = N_c-1}\, , \\ \varphi^{af} = z^f \qquad & {a = N_c}\, ,
\end{array}
\end{equation}
($z^f$ and $w^f$ are generic $N_f$-dimensional vectors) satisfy the
condition (\ref{mincond2}).  To understand which type of
configurations dominate, we have again resorted to numerical
simulations on small lattices.  The results are reported in
Table~\ref{tablargebeta}. For $\Pi_{\bm x}$ results are consistent
with
\begin{equation}
   \langle \hbox{Re Tr } \Pi_{\bm x} \rangle = 1.
\end{equation}
This relation is consistent with Eq.~(\ref{Pi-largebeta}) only if we
assume that the matrix $V$ is a randomly chosen SU($N_c-1$)
matrix. For instance, if $V = V_1 \oplus 1$ with a generic
$V_1\in$SU($N_c-2$), one would instead predict $\langle \hbox{Re Tr }
\Pi_{\bm x} \rangle = 2$.  This result constraints the field $\varphi$
to be of the form (\ref{phi-largebeta}).  If this is the case, the
operator $P$ takes the form $P^{fg} = \bar{z}^f z^g$ in the
large-$\beta$ regime.  Therefore, $P$ becomes a projector and
$\hbox{Tr }P^2$ is predicted to be one.  Analogously, also the Binder
parameter should converge to one.  The numerical results reported in
Table~\ref{tablargebeta} are in perfect agreement.

\begin{table*}
\begin{tabular}{cllllllr}
\hline\hline
$(N_c, N_f)$ & 
    \multicolumn{1}{c}{$\langle \mathrm{Re\,Tr}\,\Pi_{\bm x}\rangle/N_c$}   & 
    \multicolumn{1}{c}{$S_\phi/(2N_f)$}  &   \multicolumn{1}{c}{$U$}   &
Eq.~(\ref{LTvalue})   & $\langle \mathrm{Tr }\, P^2_{\bm x}\rangle$ & 
  Eq.~\eqref{myguess}  & $L$ \\ 
\hline
(2, 2) & 1.000018(8) & $-$1.000006(6) &1.1905(5) & 1.190476$\ldots$ &0.8000(3)&
     0.8 & 4 \\
\hline
(2, 3) & 0.999997(3) & $-$1.000002(3) &1.0937(4) & 1.093750$\ldots$ &0.7144(5)& 
0.714286$\ldots$ & 4\\ 
\hline
(3, 2) & 1.0000(1)   & $-$1.00000(1)  &1.0050(3) &&0.9795(8)& & 4 \\ 
\hline
(3, 3) & 0.99997(2)  & $-$1.000003(6) & 1.0313(7) &&   0.883(3)  &&    4 \\
(3, 3) & 0.999999(6) & $-$0.999997(3) & 1.00472(6)&&   0.9507(4) &&    6 \\
(3, 3) & 1.000007(8) & $-$1.000000(5) & 1.00100(1)&&   0.9746(1) &&    8 \\
(3, 3) & 1.000017(7) & $-$0.999999(2) & 1.00032(1)&&   0.98536(6)&&   10 \\
(3, 3) & 1.00000(2)  & $-$0.999997(2) & 1.00013(1)&&   0.9905(1) &&   12 \\
(3, 3) & 1.000015(8) & $-$0.999999(1) & 1.000045(3)&&  0.9946(2) &&   15 \\
\hline
(3, 4) & 1.0000(1)   & $-$1.000004(4) & 1.0520(5) && 0.764(2)  &&    4 \\
(3, 4) & 0.99999(1)  & $-$0.999998(3) & 1.00460(9)&& 0.9499(3) &&    6 \\
(3, 4) & 1.00001(1)  & $-$1.000000(2) & 1.00099(1)&& 0.9746(5) &&    8 \\
(3, 4) & 1.00001(1)  & $-$1.000002(3) & 1.00032(1)&& 0.98517(7)&&   10 \\
\hline
(4, 2) & 0.99998(2)  & $-$0.999998(7) &1.0014(2) &
       &0.9881(3)&  \\
\hline\hline
\end{tabular}
\caption{Asymptotic values for $\beta\to\infty$ on a $L^2$ lattice for 
$\gamma = 1$. Here $S_\phi$ is the part of the action (\ref{hgauge}) that 
depends on the $\varphi$ field (for $\gamma = 0$ we have $S_g = S_\phi$).  
}
\label{tab:largebeta-gammap}
\end{table*}

\begin{table*}
\begin{tabular}{cllllllr}
\hline\hline
$(N_c, N_f)$ & 
    \multicolumn{1}{c}{$\langle \mathrm{Re\,Tr}\,\Pi_{\bm x}\rangle/N_c$}   & 
    \multicolumn{1}{c}{$S_\phi/(2N_f)$}  &   \multicolumn{1}{c}{$U$}   &
Eq.~(\ref{LTvalue})   & $\langle \mathrm{Tr }\, P^2_{\bm x}\rangle$ & 
  Eq.~\eqref{myguess}  & $L$ \\ 
\hline
(2,2) & $-$0.80901(6)& $-$0.80898(2) & 1.190(2) & 1.190476$\ldots$ & 0.8000(5)
 & 0.8 & 4 \\
(2,2) & $-$0.80896(4)& $-$0.80903(2) & 1.1905(5)& 1.190476$\ldots$ & 0.7999(6)
 & 0.8 & 6 \\
(2,2) & $-$0.80899(3)& $-$0.809022(6)& 1.191(1) & 1.190476$\ldots$ & 0.7998(6)
 & 0.8 & 8 \\ 
\hline
(2,3) & $-$0.60988(6)& $-$0.84902(2) & 1.0938(5)& 1.093750$\ldots$ & 0.7142(6)
 & 0.714286$\ldots$ &   4 \\
(2,3) & $-$0.60982(4)& $-$0.849028(7)& 1.0937(5)& 1.093750$\ldots$ & 0.7142(6) 
 & 0.714286$\ldots$ &   6 \\
(2,3) & $-$0.60984(8)& $-$0.84902(3) & 1.0938(2)& 1.093750$\ldots$ & 0.7142(4) 
 & 0.714286$\ldots$ &   8 \\
\hline
(3,2) & $-$0.33333(2)& $-$0.999995(6)& 1.0000010(4)&&  1.00000(2) &&   4 \\
(3,2) & $-$0.333326(6)&$-$0.999998(4)& 1.0000005(2)&&  1.000000(6)&&   6 \\
(3,2) & $-$0.333343(6)&$-$0.999997(4)& 1.0000003(3)&&  1.000000(5)&&   8 \\
\hline
(3,3) & $-$0.333316(8)&$-$0.99999(2) & 1.000000(1) &&  0.99999(1) &&   4 \\
(3,3) & $-$0.333338(8)&$-$1.0000(1)  & 1.0000003(2)&&  0.99998(1) &&   6 \\
(3,3) & $-$0.333336(4)&$-$0.999995(5)& 1.0000001(1)&&  1.000004(4)&&   8 \\
\hline
(3,4) & $-$0.33331(2)& $-$0.999999(3)& 1.0000003(2)&&  0.999999(2)&&   4 \\
(3,4) & $-$0.33332(5)& $-$1.000000(3)& 1.0000001(1)&&  1.000006(5)&&   6 \\
(3,4) & $-$0.33333(1)& $-$0.999997(2)& 1.0000000(1)&&  0.999995(4)&&   8 \\
\hline\hline
\end{tabular}
\caption{Asymptotic values for $\beta\to\infty$ on a $L^2$ lattice for 
$\gamma =-1$. Here $S_\phi$ is the part of the action (\ref{hgauge}) that 
depends on the $\varphi$ field (for $\gamma = 0$ we have $S_g = S_\phi$). }
\label{tab:largebeta-gammam}
\end{table*}

\subsection{Behavior for $\gamma\not=0$}

Let us now determine the minimum-energy configurations for $\gamma >
0$.  For $N_c = 2$, the introduction of a positive $\gamma$ is of
course irrelevant: the gauge part of the action is already minimized
for $\gamma = 0$.  The analysis for $N_c \ge 3$ is instead more
tricky.  If we define $\beta_g = \beta \gamma$, fixing the value of
$\gamma$ corresponds to considering a particular way of approaching
the limiting point $\beta = \beta_g = \infty$. We will now argue that
the relevant $\varphi$ configurations, that is those that dominate the
ensemble average, strongly depend on how the limit is taken.  Imagine
that one first takes the limit $\beta_g \to \infty$ at fixed, finite
$\beta$ and then the limit $\beta\to \infty$. In this case, the
limiting configurations would have the form (\ref{minconf}) and
$\langle {\rm Tr}\, P_{\bm x} ^2 \rangle$ and $U$ would assume the
values (\ref{myguess}) and (\ref{LTvalue}), respectively.  On the
other hand, consider the opposite approach: first, we take the limit
$\beta\to\infty$ at fixed, finite $\beta_g$, followed by
$\beta_g\to\infty$. For finite $\beta_g$, the operator $\Pi_{\bm x}$
should have the form (\ref{Pi-largebeta}).  The matrix $V$ would not
be random (it should become closer to the identity as $\beta_g$
increases). Nonetheless, it is not expected to have an eigenvector
with an eigenvalue exactly equal to one. There, the relevant $\varphi$
configurations should always be of the form (\ref{phi-largebeta}).  It
is not possible to predict a priori what are the relevant
configurations if the limit is taken keeping the ratio $\gamma =
\beta_g/\beta$ fixed and we have therefore performed simulations on
very small lattices.  Results for $\gamma =1$ are reported in
Table~\ref{tab:largebeta-gammap}. For $N_f = 2$ results on a lattice
with $L=4$ are definitely consistent with $U = 1$ and $\langle
\hbox{Tr}\, P_{\bm x} ^2 \rangle=1$. For $N_f=3$ we observe
significantly larger size corrections. We have performed a detailed
study for $N_c = 3$.  We observe that the data converge to the
expected results $U = 1$ and $\langle \hbox{Tr}\, P_{\bm x} ^2
\rangle=1$.  $U$ converges quite fast, while the second quantity
converges with the expected behavior $L^{-2}$.  For $N_f = 4$, size
corrections are even larger, but the extrapolations are again
consistent with the expected results. If we extrapolate the estimates
of $\langle \hbox{Tr}\, P_{\bm x} ^2 \rangle$ reported in
Table~\ref{tab:largebeta-gammap} assuming corrections that decay as
$1/L^2$ we obtain results that are consistent with one.

Let us finally discuss the case $\gamma < 0$, which is much less
obvious, as the system shows frustration. Indeed, if we minimize the
contribution of the action that depends on the fields $\varphi$, we
obtain the consistency condition (\ref{mincond2}), which requires each
plaquette to have at least one unit eigenvalue. On the other hand,
minimizing the plaquette term we would expect the plaquette $\Pi_{\bm
  x}$ to converge to $-I$ or $-e^{\pm i\alpha} I$, for even or odd
$N_c$, respectively, where $I$ is the $N_c\times N_c$ identity matrix
and $\alpha = \pi/N_c$ [correspondingly, $\hbox{Tr}\, \Pi_{\bm x}$
  would converge to $-N_c$ or $-N_c \cos (\pi/N_c)$]. Therefore, one
cannot simultaneously minimize all local contributions: the system is
frustrated.

To understand the effective behavior along the lines $\beta_g = \gamma
\beta$, we have again studied numerically the system for a specific
value of $\gamma$, $\gamma=-1$.  The results are reported in
Table~\ref{tab:largebeta-gammam}.  Per $N_c = 2$, they show the clear
presence of frustration. The plaquette is not identical to the matrix
$-I$ and the minimal condition (\ref{minenea}) is not
satisfied. Nonetheless, the estimates of the Binder parameter and of
the trace of $P^2_{\bm x}$ are completely consistent with the the
results obtained for $\gamma = 0$. Even for $\gamma=-1$ the fields
$\varphi$ are uniformly distributed on the $N$-dimensional sphere ($N
= 4 N_f$). We conclude that $\gamma$ plays a role only on the gauge
properties, but not on the behavior of $\varphi$ correlations, which
dominate the large-$\beta$ behavior.

For $N_c=3$, the results for the trace of the plaquette are completely
consistent with $\Pi_{\bm x} = \hbox{diag}\, (-1,-1,1)$. In other
words, the relevant gauge configurations have the same structure that
holds for $\gamma = 0$, Eq.~(\ref{Pi-largebeta}). The only difference
is that the matrix $V$ is no longer a random SU(2) matrix, but is
simply $-I$ (where $I$ is the two-dimensional identity). Finally, we
have some results for $N_c=4$. In this case, it is difficult to take
the limit $\beta \to \infty$, using data in the range $10\le \beta\le
150$ as data still significantly change as $\beta$ increases.  For
$P^2$ and $U$ results are close to the expected ones. For a system of
size $L=4$, we obtain for the average trace of $P^2_{\bm x}$, 0.951
and 0.984 for $N_f=2$ and $\beta = 50, 150$, respectively (statistical
errors are very small, of order $10^{-5}$); for $N_f = 5$, we obtain
instead 0.931 and 0.978 again for $\beta = 50, 150$. Again, results
are consistent with a limiting value of one.


\begin{thebibliography}{99}

\bibitem{Weinberg-book} S. Weinberg, {\em The Quantum Theory of
  Fields}, (Cambridge University Press, 2005).

\bibitem{WNMXS-17} C. Wang, A. Nahum, M. A. Metlitski, C. Xu, and
  T. Senthil, Deconfined Quantum Critical Points: Symmetries and
  Dualities, Phys. Rev. X {\bf 7}, 031051 (2017).

\bibitem{GASVW-18} S. Gazit, F. F. Assaad, S. Sachdev, A. Vishwanath,
  and C. Wang, Confinement transition of $Z_2$ gauge theories coupled
  to massless fermions: emergent QCD$_3$ and SO(5) symmetry,
  Proceedings of the National Academy of Sciences 115, E6987 (2018).

\bibitem{SSST-19} S. Sachdev, H. D. Scammell, M. S. Scheurer, and
  G. Tarnopolsky, Gauge theory for the cuprates near optimal doping,
  Phys. Rev. B {\bf 99}, 165126 (2019).

\bibitem{GSF-19} H. Goldman, R. Sohal, and E. Fradkin, Landau-Ginzburg
  Theories of non-Abelian quantum Hall states from non-Abelian
  bosonization, Phys. Rev. B {\bf 100}, 115111 (2019).

\bibitem{Sachdev-19} S. Sachdev, Topological order, emergent gauge
  fields, and Fermi surface reconstruction, Rep. Prog. Phys. {\bf 82},
  014001 (2019).

\bibitem{BPV-19} C. Bonati, A. Pelissetto and E. Vicari, Phase
  diagram, symmetry breaking, and critical behavior of
  three-dimensional lattice multiflavor scalar chromodynamics,
  Phys. Rev. Lett. {\bf 123}, 232002 (2019).

\bibitem{BPV-20} C. Bonati, A. Pelissetto and E. Vicari,
  Three-dimensional lattice multiflavor scalar chromodynamics:
  interplay between global and gauge symmetries, 
  arXiv:2001.01132.


\bibitem{MW-66} N.~D.~Mermin and H.~Wagner, Absence of ferromagnetism
  or antiferromagnetism in one- or two-dimensional isotropic
  Heisenberg models, Phys. Rev. Lett.  {\bf 17}, 1133 (1966).

\bibitem{BHZ-80} E. Br\'ezin, S. Hikami, and J. Zinn-Justin,
  Generalized non-linear $\sigma$-models with gauge invariance,
  Nucl. Phys. B {\bf 165}, 528 (1980).

\bibitem{ZJ-book} J. Zinn-Justin, 
  {\em Quantum Field Theory and Critical Phenomena}, 
  fourth edition (Clarendon Press, Oxford, 2002).

\bibitem{Wilson-74} K.G. Wilson, Confinement of quarks, Phys. Rev. D
  {\bf 10}, 2445 (1974).

\bibitem{Georgi-book} H.~Georgi, {\em Weak interactions and modern
  particle theory}, Benjamin/Cummings Publishing Company (1984).

\bibitem{DP-14} P.~S.~Bhupal Dev and A.~Pilaftsis, Maximally symmetric
  two Higgs doublet model with natural standard model alignment, JHEP
  {\bf 1412}, 024 (2014); Erratum: [JHEP {\bf 1511}, 147 (2015)]


\bibitem{FB-72} M. E. Fisher and M. N. Barber, Scaling theory for
  finite-size effects in the critical region, Phys. Rev. Lett. {\bf
    28}, 1516 (1972).

\bibitem{Barber-83}
  M. N. Barber, in {\em Phase Transitions and Critical Phenomena},
  edited by C. Domb and J. L. Lebowitz (Academic Press, New York, 1983),
  Vol. 8.

\bibitem{Privman-90} V. Privman ed., {\em Finite Size Scaling and
  Numerical Simulation of Statistical Systems} \/ (World Scientific,
  Singapore, 1990).

\bibitem{PV-02} A. Pelissetto and E. Vicari, Critical phenomena and
  renormalization group theory, Phys. Rep. {\bf 368}, 549 (2002).

\bibitem{BPV-19-2} C. Bonati, A. Pelissetto and E. Vicari,
  Two-dimensional multicomponent Abelian-Higgs lattice models,
  arXiv:1912.01315.

\bibitem{CP-98} 
S. Caracciolo and A. Pelissetto, Corrections to
  finite-size scaling in the lattice $N$-vector model for $N=\infty$,
  Phys. Rev. D {\bf 58}, 105007 (1998).

\bibitem{KT-73} 
J. M. Kosterlitz and  D. J. Thouless,
  Ordering, metastability and phase transitions in two-dimensional systems,
  J.\ Phys. C: Solid State {\bf 6},  1181 (1973).

\bibitem{Berezinskii-70} V. L. Berezinskii, Destruction of Long-range
  Order in One- dimensional and Two-dimensional Systems having a
  Continuous Symmetry Group I. Classical Systems,
  Zh. Eksp. Theor. Fiz. {\bf 59}, 907 (1970) [Sov. Phys. JETP {\bf
      32}, 493 (1971)].

\bibitem{Kosterlitz-74}
  J. M. Kosterlitz, 
  The critical properties of the two- dimensional xy model,
  J. Phys. C {\bf 7}, 1046 (1974).

\bibitem{JKKN-77} J. V. Jos\'e, L. P. Kadanoff, S. Kirkpatrick, and
  D. R. Nelson, Renormalization, vortices, and symmetry-breaking
  perturbations in the two-dimensional planar model, Phys. Rev. B
  {\bf 16}, 1217 (1977).

\bibitem{Witten-79} E. Witten, Instantons, the quark model, and the
  $1/N$ expansion, Nucl. Phys. B {\bf 149}, 285 (1979).

\bibitem{CR-93} M. Campostrini and P. Rossi, The $1/N$ expansion of
  two-dimensional spin models, Riv. Nuovo Cimento {\bf 16}, 1 (1993).

\bibitem{CRV-92} M. Campostrini, P. Rossi, and E. Vicari, Monte Carlo
  simulation of CP$^{N-1}$ models, Phys. Rev. D {\bf 46}, 2647 (1992).


\bibitem{footnote-interpolation} An interpolation of the MC data of
  $U$ versus $R_\xi$ for the lattice 2D O(5) vector model is provided
  by the function $f(y) = 1.40000 -(40.78626 + 232.69415\, y +
  626.04078 \, y^2)\,h(y) +266.55101 \,y^2 -116.42105 \,y^3
  -98.87100\, y^4 + 73.23167 \,y^5 - 10.81759\, y^6$, where $h(y) = 2
  y (1 - e^{-10 y})/(1 + 2 e^{2y})$.  This is valid for $y\in
  [0,0.8]$. The error is estimated to be smaller than 0.5\%, including
  the uncertainty due to the scaling corrections.


\bibitem{PV-19} A. Pelissetto and E. Vicari, Multicomponent compact
  abelian Higgs models, Phys. Rev E {\bf 100}, 042134 (2019);
  Three-dimensional ferromagnetic CP$^{N-1}$ models, Phys. Rev. E {\bf
  100}, 022122 (2019).

\bibitem{PRV-01}
  A. Pelissetto, P. Rossi, and E. Vicari,
  Large-$N$ critical behavior of O($M$)$\times$O($N$) spin models, 
  Nucl. Phys. B {\bf 607}, 605 (2001).

\end{thebibliography}
\end{document}